\documentclass[twocolumn,superscriptaddress,prb,10pt]{revtex4-1}
\usepackage{amsmath,amssymb}
\usepackage{graphicx}
\usepackage{xcolor}
\usepackage[colorlinks,bookmarks=false,citecolor=blue,linkcolor=red,urlcolor=blue]{hyperref}
\usepackage{subfigure}

\begin{document}

\title{The entanglement negativity in random spin chains} 

\author{Paola Ruggiero}
\affiliation{International School for Advanced Studies (SISSA),
Via Bonomea 265, 34136, Trieste, Italy, 
INFN, Sezione di Trieste}

\author{Vincenzo Alba}
\affiliation{International School for Advanced Studies (SISSA),
Via Bonomea 265, 34136, Trieste, Italy, 
INFN, Sezione di Trieste}

\author{Pasquale Calabrese}
\affiliation{International School for Advanced Studies (SISSA),
Via Bonomea 265, 34136, Trieste, Italy, 
INFN, Sezione di Trieste}

\date{\today}

\begin{abstract} 

We investigate the logarithmic negativity in strongly-disordered spin chains in the random-singlet phase. 
We focus on the spin-$1/2$ random Heisenberg chain and the random $XX$ chain. 
We find that for two arbitrary intervals the disorder-averaged negativity and the mutual information are proportional 
to the number of singlets shared between the two intervals. 
Using the strong-disorder renormalization group (SDRG), 
we prove that the negativity of two adjacent intervals grows logarithmically with the intervals length. 
In particular, the scaling behavior is the same as in  conformal field theory, but with a different prefactor.  
For two disjoint intervals the negativity is given by a universal simple function of the cross ratio, reflecting scale invariance. 
As a function of the distance of the two intervals, the negativity  decays algebraically in contrast with the  
exponential behavior in clean models. 
We confirm our predictions using a numerical implementation of the SDRG method.  
Finally, we also implement DMRG simulations for the negativity in open spin chains. The chains accessible in the 
presence of strong disorder are not sufficiently long to provide a reliable confirmation of the SDRG results. 

\end{abstract}


\maketitle

\section{Introduction}
\label{intro}

Entanglement measures are nowadays accepted as powerful tools to characterize 
quantum many-body systems~\cite{amico-2008,calabrese-2009,eisert-2010,laflorencie-2016}. 
Arguably, the most popular and useful one is the entanglement entropy. 
Given a {\it pure} state $|\psi\rangle$ and a bipartition into two parts $A$ and $B$, 
the entanglement entropy of part $A$ 
is defined as 
\begin{equation} 
S_A\equiv-\textrm{Tr}\rho_A\ln\rho_A,
\end{equation}
where $\rho_A\equiv\textrm{Tr}_B|\psi\rangle\langle\psi|$ is the reduced density matrix of $A$. 
For a pure state it is clear that $S_A=S_B$, reflecting the property that a good measure of entanglement is symmetric in $A$ and $B$.

\begin{figure}[b]
\includegraphics[width=0.48\textwidth]{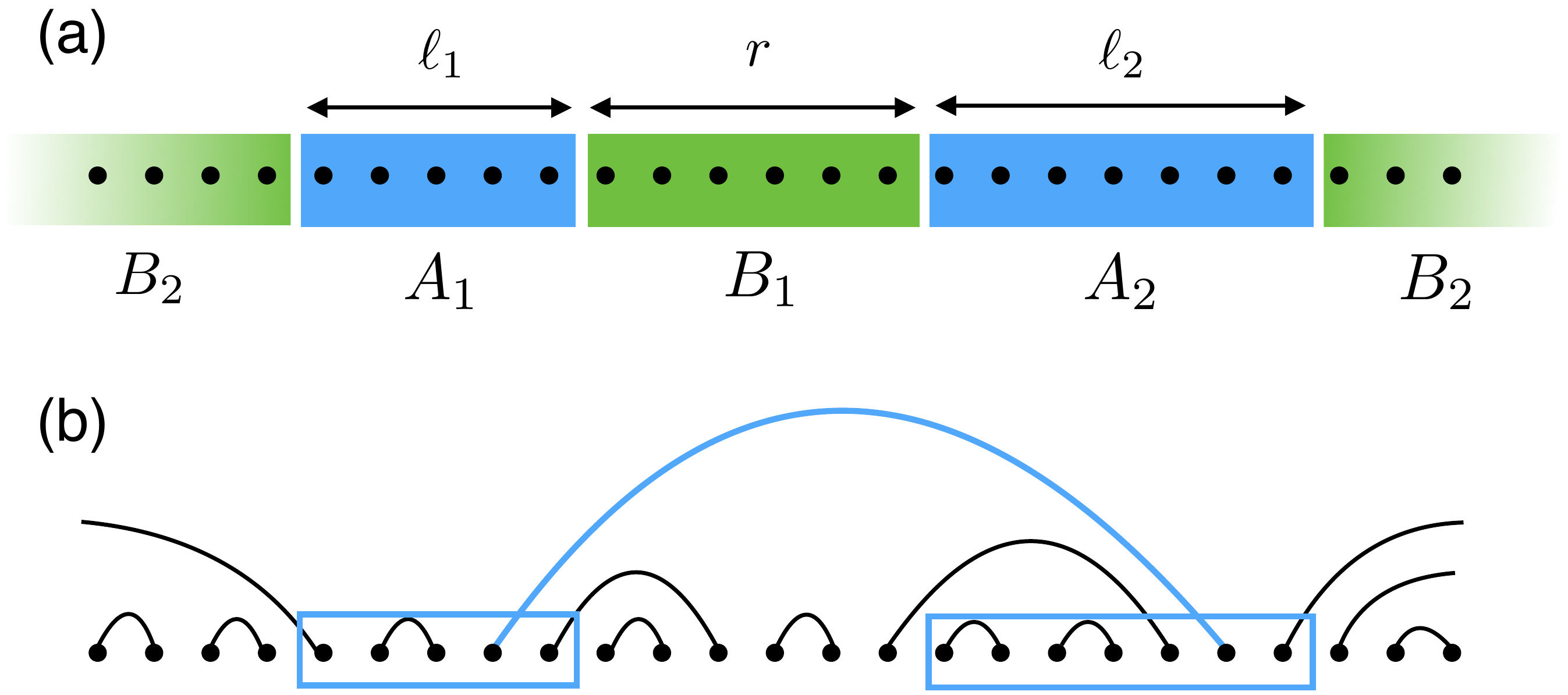}
\caption{(\textcolor{red}{a}): Partition of an infinite spin chain corresponding
 to two disjoint intervals. The region of interest is $A_1\cup A_2$. 
 Two adjacent intervals correspond to $r=0$. (\textcolor{red}{b}): 
 Cartoon of the random singlet phase. The lines connect pairs of spins forming $SU(2)$ singlets. 
 The entanglement negativity ${\cal E}_{A_1:A_2}$ is proportional  
 to the number of singlets $n_{A_1:A_2}$ shared between $A_1$ and $A_2$ 
 (here $n_{A_1 : A_2}= 1$). 
}
\label{figure2}
\end{figure}

If a system is in a {\it mixed} state, for instance at finite temperature, 
a useful measure of the correlation between $A$ and $B$ is the mutual information 
${\mathcal I}_{A:B}$, which is defined as the symmetrized combination ${\mathcal I
}_{A:B}\equiv S_A+S_B-S_{A\cup B}$. However, it is well known that the mutual 
information provides only an upper bound for the entanglement between $A$ and $B$, 
as it is sensitive to both classical and quantum correlations. A similar issue arises 
when quantifying the entanglement between disconnected regions in pure states. 
For instance, given the tripartition of a system as $B_2\cup A_1\cup B_1\cup A_2\cup B_2$ 
(as illustrated in Fig.~\ref{figure2} (a) for a spin chain), with $A\equiv A_1\cup A_2$ the 
region  of interest,  ${\mathcal I}_{A_1:A_2}$ is not a measure of the mutual entanglement 
between $A_1$ and $A_2$. 

A computable measure of the mutual entanglement between two subsystems in a mixed state is provided 
by the so-called {\it logarithmic negativity}~\cite{peres-1996,zycz-1998,zycz-1999,lee-2000,
vidal-2002,plenio-2005} 
\begin{equation}
{\cal E}\equiv \ln||\rho_A^{T_2}||_1=\ln\textrm{Tr}|\rho^{T_2}_A|. 
\end{equation}
Here $\rho_A^{T_2}$ is the partially transposed reduced density matrix with respect 
to $A_2$. This is defined as $\langle\varphi_1\varphi_2|\rho^{T_2}_A|\varphi'_1
\varphi'_2\rangle\equiv\langle\varphi_1\varphi_2'|\rho_A|\varphi'_1\varphi_2\rangle$, 
with $\{\varphi_1\}$ and $\{\varphi_2\}$ being bases for $A_1$ and $A_2$, respectively. 

Besides its interest in quantum information, recently the negativity became the focus of 
several interdisciplinary efforts to understand its role in many-body systems. 
For critical points described by a Conformal Field Theory 
(CFT) its scaling behavior has been derived analytically~\cite{calabrese-2012,cct-neg-long,calabrese-2013}. 
This allowed to prove that, unlike the entanglement entropy, the negativity is scale invariant 
in  gapless systems described by CFT, as it was already argued based on DMRG~\cite{white-1992,uli1,uli2} 
simulations~\cite{hannu-2008}, 
on semi-analytic results for the harmonic chain \cite{mrpr-09}, and for mean-field exactly solvable spin models~\cite{hannu-2010}. 
Furthermore, 
for disjoint intervals the logarithmic negativity contains, in principle, complete information 
about the intricate structure of the underlying CFT~\cite{calabrese-2012} (as it does the entanglement entropy of two or more
disjoint intervals \cite{2int}). 
Some of these results have been generalized to finite-temperature~\cite{calabrese-2015}, 
large central charge \cite{kpp-14},
out of equilibrium situations \cite{ctc-14,hoogeveen-2015,eisler-2014,wen-2015}, and holographic\cite{rr-15} and 
massive quantum field theories~\cite{fournier-2015}. 
In topologically ordered phases, i.e., characterized by a non-zero topological entropy~\cite{levin-2006,kitaev-2006}, it has been shown 
that the negativity is sensitive  only to the ``quantum'' contribution~\cite{lee-2013,
castelnovo-2013}, while it is zero in classical topologically ordered states. Also, 
the negativity proved to be a useful tool to characterize Kondo-like behavior in spin 
chains~\cite{bayat-2012,bayat-2014,abab-16}. Surprisingly, the exact treatment of the partial 
transposition in free fermion models is an arduous task, and no exact results are known 
yet for the negativity, despite some recent progresses for the calculations of the moments of 
the partial transpose~\cite{eisler-2014,coser-2015,ctc-16,ctc-16b,chang-2016,hw-16}. 
Oppositely, the negativity can be calculated analytically for free bosonic models~\cite{audenaert-2002}. 
for which few results are also available in higher dimensions~\cite{eisler-2016,dct-16}. 
Furthermore, some interesting results on the negativity in Chern-Simons 
theories have been provided recently~\cite{wen-2016}. Finally, from the numerical 
perspective, effective methods to calculate the negativity are available in the context 
of Tree Tensor Networks~\cite{calabrese-2013}, while the moments of the partially 
transposed reduced density matrix can be calculated using classical~\cite{alba-2013} 
and quantum~\cite{chung-2014} Monte Carlo techniques. 
When not analytically possible, numerical extrapolations can be used to obtain the negativity from the replica limit 
of the moments \cite{dct-15}.

At the same time the study of the interplay between disorder and entanglement became a fruitful research 
area~\cite{refael-2009,laflorencie-2016}. 
For instance, in Ref.~\onlinecite{RefaelMoore2004} it has been shown that, for 
disordered spin chains exhibiting the random singlet (RS) phase, the scaling of the 
disorder averaged entanglement entropy is logarithmic with the subsystem size like in a CFT. 
This has been tested numerically in the random $XX$ chain\cite{laflorencie-2005}
(which is exactly solvable for each realization of the disorder), and in the random $XXX$ chain~\cite{dechiara-2006} 
using DMRG. 
Furthermore, the moments of the reduced density matrix $\textrm{Tr}\rho_A^\alpha$ 
have been also studied~\cite{FagottiCalabreseMoore2011}, as well as the spectrum \cite{py-13}, 
and the entanglement in low-lying excited states \cite{ramirez-2014}.
Other disordered spin 
models \cite{raul-2006,refael-2007,hvlm-07,lir-07,by-07,bcmr-07,il-08,frbm-08,ysh-08,hlvv-11,ki-12,gah-16,rsrs-16} have been 
also considered, obtaining  similar results for the scaling of the entanglement entropy. 
The non-equilibrium features of the entanglement in these random spin chains are also under intensive investigation 
\cite{dechiara-2006,bo-07,isl-12,bpm-12,spa-13,va-14,prado-14,vm-15,zas-16}.
The behavior of entanglement related quantities in {\it classical} disordered spin systems has been explored~\cite{alba-2016}. 

In this paper we investigate the disorder averaged logarithmic negativity in RS phases 
in the framework of the strong-disorder renormalization group (SDRG)~\cite{Igloi-rev}. 
We focus on the spin-$1/2$ Heisenberg ($XXX$) chain with random antiferromagnetic couplings, 
and on the random $XX$ chain. 
We consider both adjacent and disjoint intervals (see Fig.~\ref{figure2} (a)). 
We demonstrate that in a RS phase the negativity of two intervals is always proportional 
to the number of singlets shared between them. 
Surprisingly, due to the simple structure 
of the RS phase, this is also the case for the mutual information, which is given as ${\mathcal 
I}_{A_1:A_2}=2{\cal E}_{A_1:A_2}$. 
More quantitatively, we find that for two adjacent intervals embedded in an infinite 
chain the disorder-averaged negativity scales as 
\begin{equation}
\label{adj-intro}
{\cal E}_{A_1:A_2}=\frac{\ln 2}{6}\ln\Big(\frac{\ell_1\ell_2}{\ell_1+\ell_2}\Big)+k, 
\end{equation}
for large $\ell_1$ and $\ell_2$ (the lengths of the two intervals). 
Here  $k$ is an additive constant. 
Interestingly, Eq.~\eqref{adj-intro} has the same functional dependence on $\ell_{1,2}$ as in 
CFT~\cite{calabrese-2012}, but the prefactor of the logarithm is different. 
For two disjoint intervals (see Fig.~\ref{figure2} (a)), we obtain 
\begin{equation}
\label{dis-intro}
{\cal E}_{A_1:A_2}=\frac{\ln2}{6}\ln \frac{(\ell_1+r)(\ell_2+r)}{r(\ell_1+\ell_2+r)}=
-\frac{\ln2}{6}\ln (1-x). 
\end{equation}
Here $\ell_1,\ell_2$ are the intervals length, and $r$ their distance. 
Eq.~\eqref{dis-intro} is 
{\it scale invariant} and for this reason the r.h.s. has been written as a function of the cross ratio 
\begin{equation}
x=\frac{\ell_1\ell_2}{(\ell_1+r)(\ell_1+r)}.
\label{xdef}
\end{equation} 
Again, this is similar to the CFT case, where 
${\cal E}_{A_1:A_2}$ depends in a more intricate way on $x$~\cite{calabrese-2012,cct-neg-long}. 
Interestingly, in the limit of two intervals far apart, from~\eqref{dis-intro} one has the power-law decay ${\cal E}\propto r^{-2}$, in 
contrast with the CFT case, where this decay is exponential~\cite{calabrese-2012}. Both~\eqref{adj-intro} 
and~\eqref{dis-intro} turn out to be in perfect agreement with exact results obtained from a numerical 
implementation of the SDRG for the random $XXX$ chain.

Finally, by generalizing the method of Ref.~\onlinecite{hannu-2008}, we discuss how to obtain 
the logarithmic negativity in DMRG simulations~\cite{itensor} 
for arbitrary tripartitions of the spin chain.  
The computational cost of the algorithm is $\propto\chi^6_{max}$, with $\chi_{max}$ the maximal bond 
dimension of the matrix product state (MPS). As a byproduct of our analysis we numerically verify the CFT scaling of the 
negativity of two adjacent intervals in the clean $XX$ chain.
In the presence of randomness it is numerically  
challenging to ensure the convergence of DMRG, introducing a systematic error in the numerical 
data. The latter is negligible only for small chains and weak disorder. The accessible chain 
sizes are not sufficient to provide reliable numerical evidence for~\eqref{adj-intro} 
and~\eqref{dis-intro}. 

The manuscript is organized as follows. In section~\ref{the-model} we introduce the $XXZ$ spin-$1/2$ 
chain with random antiferromagnetic couplings and  summarize the strong 
disorder renormalization group (SDRG) method. 
In section~\ref{SECTIONnegativityRSP} we provide the 
analytic expressions for the logarithmic negativity and the moments of the partially transposed reduced density matrix
in RS phases. 
The scaling of the negativity is discussed in section~\ref{scal-neg} for both adjacent and disjoint 
intervals. These results are verified in section~\ref{n-sdrg} using a numerical implemtation 
of the SDRG method. Section~\ref{neg-dmrg} focuses on the calculation of the logarithmic negativity 
in DMRG simulations. 
In section~\ref{neg-dmrg1} 
we present DMRG results for the negativity of two adjacent intervals in the random 
$XX$ chain. Finally, we conclude in section~\ref{concl}.

\section{The disordered Heisenberg spin chain and the strong disorder RG}
\label{the-model}

The random antiferromagnetic spin-$\frac{1}{2}$ $XXZ$ chain with open boundary 
conditions (OBC) is defined by the Hamiltonian 
\begin{equation}
{\mathcal H}=\sum\limits_{i=1}^{L-1}J_i(S^x_iS^x_{i+1}+S^y_i
S^y_{i+1} + \Delta S^z_i
S^z_{i+1}  ), 
\label{xxz-ham}
\end{equation}
where $S_i^{x,y,z}$ are the spin components acting on site $i$, $L$ is the length of the chain, $\Delta$ 
the anisotropy parameter, and $\{J_i\}_{i=1}^{L-1}$ are uncorrelated 
 {\it positive} random variables, drawn from a distribution $P(J)$. 
For periodic boundary conditions (PBC) one has an extra term in~\eqref{xxz-ham} 
connecting site $L$ with site $1$. We focus on $\Delta=1$ and $\Delta=0$, corresponding 
to the $XXX$ and the $XX$ random chain, respectively. For generic $\{J_i\}_{i=
1}^{L-1}$, the latter can be treated analytically for each realization of the disorder, 
exploiting the mapping to  free fermions (see Appendix~\ref{dis-XX}). 
Here we restrict ourselves to the family of 
distributions 
\begin{equation}
P_\delta(J)\equiv\delta^{-1}J^{-1+1/\delta},
\end{equation}
with $J\in [0,1]$, and 
$0\le\delta<\infty$ a parameter tuning the disorder strength. For $\delta\to0$ one 
recovers the clean, i.e., without disorder, $XXZ$ chain, whereas $\delta\to\infty$ 
corresponds to the infinite-randomness fixed point (IRFP, see section~\ref{the-model}). 
The latter describes the low-energy physics of~\eqref{xxz-ham},  irrespective of the 
chosen distribution $P(J)$. For $\delta=1$, $P_\delta(J)$ becomes the flat distribution 
(box distribution) in the interval $[0,1]$.


We now briefly review the SDRG method for the $XXX$ chain. The main idea is to 
obtain a low-energy effective description of~\eqref{xxz-ham} by successively 
integrating out the strongest couplings, and renormalizing the remaining ones. 
Given an arbitrary coupling configuration $\{J_i\}_{i=1}^{L-1}$, one starts by  
identifying the strongest bond $J_M\equiv\textrm{max}_i\{J_i\}$. The interaction 
between the two spins coupled by $J_M$ (that we denote as $\vec S_l$ and $\vec 
S_r$) is given by the Hamiltonian ${\mathcal H}_0$ as 
\begin{equation}
{\mathcal H}_0 = J_M \vec{S}_l \cdot \vec{S}_r. 
\end{equation}
The ground state of ${\mathcal H}_0$ is the singlet state $|s\rangle$
\begin{equation}
|s\rangle\equiv2^{-1/2}(\left|\uparrow_l\downarrow_r\right\rangle  - \left|
\downarrow_l \uparrow_r\right\rangle). 
\end{equation}
The interaction between $\vec S_l$ and $\vec S_r$, and their neighboring spins (here 
denoted as $S'_l$ and $S'_r$, respectively) is described by the Hamiltonian 
${\mathcal H}'$ as 
\begin{equation}
{\mathcal H}' = J_l \; \vec{S}'_l \cdot \vec{S}_{l} + J_r \; \vec{S}_r \cdot 
\vec{S'}_r. 
\end{equation}
Since by definition $J_l,J_r<J_M$, one can treat ${\mathcal H}'$ as a 
perturbation. Within second-order perturbation theory, this leads to the 
effective Hamiltonian ${\mathcal H}^{eff}$ for $\vec S_l',\vec S_r'$ as 
\begin{multline}
{\mathcal H}^{eff} = \langle s | {\mathcal H}_0 + {\mathcal H}'|s\rangle+
\sum_{t} \frac{ |\langle s | {\mathcal H}' | t  \rangle |^2 }{E_s - E_t}\\
\equiv E_0 + J' \; \vec{S}'_l \cdot \vec{S}'_r. 
\label{perturbation}
\end{multline}
Here the sum is over the triplet states of two spins $|t\rangle=\left|
\uparrow\uparrow\right\rangle,(\left|\uparrow\downarrow\right\rangle+\left
|\downarrow\uparrow\right\rangle)/\sqrt{2},\left|\downarrow\downarrow\right
\rangle$. The corresponding energies are $E_{t}\equiv\langle t|{\mathcal H}_0
|t\rangle=1/4J_M$, whereas one has $E_s\equiv\langle s|{\mathcal 
H}_0|s\rangle=-3/4J_M$. In the last step in~\eqref{perturbation} we defined 
$E_0= -3(4J_M + 3J_l^2 + 3J_r^2)/16$. The effective coupling $J'$ between 
$\vec S_l'$ and $\vec S_r'$ reads 
\begin{equation}
\label{ren-coup}
J'= \frac{J_l J_r}{2J_M}. 
\end{equation}
Note that~(\ref{perturbation}) does not depend on $\vec S_l, \vec S_r$ anymore. 
Moreover, ${\mathcal H}^{eff}$ is still of the Heisenberg form (\ref{xxz-ham}) with the 
renormalized coupling $J'$. This process of decimating the spins connected 
by the strongest bond, renormalizing the remaining interactions, can be 
represented as  
\begin{equation}
\label{DMrule}
\Big(\cdots, J_l , J_M , J_r, \cdots\Big)_L \; \to \; \Big(\cdots , 
\frac{J_l J_r}{2 J_M}, \cdots\Big)_{L-2}, 
\end{equation}
and it defines the so-called Dasgupta-Ma rule~\cite{DasguptaMa}. Crucially, 
since $J_l,J_r<J_M$, the repeated application of~\eqref{DMrule} reduces the energy 
scale of the model. 

The low-energy properties of the model are asymptotically, i.e., after many 
iterations of~\eqref{DMrule}, described by the so-called \emph{random singlet} 
(RS) phase. This is illustrated in Fig.~\ref{figure2} (b). In the RS phase all the 
spins are paired (as stressed by the links in the figure) in a random 
fashion. Paired spins form singlet. From~\eqref{DMrule} it is clear that longer 
range singlets are generated at later steps of the SDRG. 


The RS phase can be quantitatively characterized through the asymptotic distribution 
of the couplings $\{J_i\}$. The iteration of the Dasgupta-Ma rule~\eqref{DMrule} leads 
to a flow for $P(J)$. It is convenient to introduce at the given SDRG step $m$ the 
variables $\beta_i^{(m)}$ and $\Gamma^{(m)}$ as 
\begin{equation}
\label{beta_Gamma}
\beta_i^{(m)} \equiv \ln \frac{J_M^{(m)}}{J_i^{(m)}}, 
\quad \Gamma^{(m)} \equiv \ln  \frac{J_M^{(0)}}{J_M^{(m)}}. 
\end{equation}
Here $J_M^{(m)}$ is the maximum coupling at the step $m$. The physical interpretation is that 
$\Gamma^{{(m)}}$ quantifies the difference in energy scales between the 
initial step and step $m$, while the  $\beta_i$ measure the broadness of the energy scales  at a fixed step $m$. 
The equation describing the SDRG flow 
of $P(J)$ is given as~\cite{ReviewIgloi}
\begin{multline}
\frac{d P}{d \Gamma} =  \frac{\partial P (\beta)}{\partial 
\beta} P(0) \; \times \\
\times  \int_0^{\infty} d \beta_1 \int_0^{\infty} d 
\beta_2 \delta (\beta - \beta_1 - \beta_2) P (\beta_1) P (\beta_2).
\label{sdrg-flow}
\end{multline}
It can be shown that~\eqref{sdrg-flow} has a unique solution $P^*(\beta)$ 
given by
\begin{equation} 
\label{Pfixedpoint}
P^* (\beta) = \frac{1}{\Gamma}\exp\Big({- \frac{\beta}{\Gamma}}\Big).
\end{equation}
Here $P^*(\beta)$ represents the fixed point of the SDRG flow. This fixed point is 
known as \emph{infinite randomness fixed point} (IRFP)~\cite{ReviewIgloi}, 
to emphasize that the broadness of the distribution
increases during the flow. 
This is reflected in $P^*(\beta)$ being flat, which corresponds to $P^*(J)$ 
being peaked at $J=0$ (cf.~\eqref{beta_Gamma}). Note that this justifies the perturbative 
treatment~\cite{Fisher1994} of ${\mathcal H}'$ in~\eqref{perturbation}. 
Remarkably,~\eqref{Pfixedpoint} does not depend on the initial distribution which is a manifestation of its universality. 

Finally, we remind the reader that for the  $XX$ chain ($\Delta= 0$ in~\eqref{xxz-ham}) the Dasgupta-Ma 
rule \eqref{DMrule} has to be modified as 
\begin{equation}
\label{DMruleXX}
\Big(\cdots, J_l , J_M , J_r, \cdots\Big)_L \; \to \; \Big(\cdots , 
\frac{J_l J_r}{J_M}, \cdots\Big)_{L-2}.
\end{equation}
Consequently, the flow equation~\eqref{sdrg-flow}, and the fixed point distribution~\eqref{Pfixedpoint} remain the same as in the
XXX case.

\section{Logarithmic negativity in random singlet phases} 
\label{SECTIONnegativityRSP}

For a generic realization of the disorder, assuming that the chain is in a RS phase, 
the logarithmic negativity between two subsystems of the chain  can be calculated analytically. 
In this section we show that it is proportional to the number of singlets shared between the two intervals. 
Let us consider a partition of the chain as in 
Fig.~\ref{figure2} with $A_1\cup A_2$ the two intervals of interest and $B_1\cup 
B_2$ their complement. For later convenience let us define
\begin{equation}
\label{part-def}
A= A_1 \cup A_2, \qquad B = B_1 \cup B_2.
\end{equation}
Given any two blocks $X,Y$ in the chain, we denote as $n_{X:Y}$ the number of singlets 
shared between them. 

Before considering the entire RS phase, it is instructive to write down the density matrix and its partial transpose
for an isolated singlet (more generic situations have been also considered in the 
literature~\cite{kor}). 
The density matrix $\rho_{2S}$ of two spins forming a singlet is 
\begin{equation} 
\label{rho2S}
\rho_{2S} =
\frac{1}{2}\begin{pmatrix}
0 & 0   &  0 & 0\\
0 & 1   & -1 & 0\\
0 & -1  &  1 & 0\\
0 & 0   &  0 & 0
\end{pmatrix},
\end{equation}
in the basis $\left|\uparrow\uparrow\right\rangle$, $\left|\uparrow\downarrow
\right\rangle$, $\left|\downarrow\uparrow\right\rangle$, and $\left|\downarrow
\downarrow\right\rangle$. The reduced density matrix $\rho_S$ for one of the spins is 
\begin{equation}
\rho_S =
\frac{1}{2}\begin{pmatrix}
1  & 0 \\
0 &  1  
\end{pmatrix}.
\end{equation}
In order to calculate the negativity, we need  the partial transpose $\rho^{T_2}_{2S}$ that is  
\begin{equation}
\rho^{T_2}_{2S}=
\frac{1}{2}
\begin{pmatrix}
 0 & 0  & 0 & -1 \\
 0 & 1  & 0 & 0 \\
 0 & 0  & 1 & 0 \\
-1 & 0  & 0 & 0 
\end{pmatrix},
\label{rhot2}
\end{equation}
with eigenvalues $\{-1/2,1/2,1/2,1/2\}$.

The density matrix $\rho_{RSP}$ for a chain in a random singlet phase is 
the tensor product of the density matrices of its constituent singlets. 
Thus, for the partition in Fig.~\ref{figure2},  one can write 
\begin{equation}
\label{ff1}
\rho_{RSP}  = \bigotimes_{i=1}^{n_{A: A}} \rho_{2S} 
\bigotimes_{i=1}^{n_{B:B}} \rho_{2S} \bigotimes_{i=1}^{n_{A:B}} \rho_{2S}. 
\end{equation}
The reduced density matrix $\rho_A$ is obtained from~\eqref{ff1} as 
\begin{equation}
\label{rhoA}
\rho_A= \bigotimes_{i=1}^{n_{A:A}} \rho_{2S}\bigotimes_{i=1}^{n_{A: B}} \rho_{S}, 
\end{equation}
and its partial transpose 
\begin{equation}
\label{rhoAT}
\rho_A^{T_2}= \bigotimes_{i=1}^{n_{A:A}} \rho_{2S}^{T_2} \bigotimes_{i=1}^{n_{A: B}} \rho_{S}^{T_2}. 
\end{equation}

To compute the negativity ${\cal E}_{A_1:A_2}$ one can  exploit the 
additivity of the negativity on tensor products, i.e., that $\mathcal{E} 
\big(\otimes_i \rho_i \big) =\sum_i \mathcal{E} (\rho_i )$. From~\eqref{rhoAT}  
one then has 
\begin{multline}
\label{neg}
\nonumber
\mathcal{E}_{A_1 : A_2} = n_{A_1:A_2}\ln\textrm{Tr}\big|\rho^{T_2}_{2S}
(A_1\cup A_2)\big|\\+\sum_{i=1,2}\Big[{n_{A_i :A_i}}\ln\textrm{Tr}
\big|\rho_{2S}^{T_2}(A_i)\big|+ n_{A_i:B}\ln\textrm{Tr}\big|\rho^{T_2}_{S}(A_i)
\big|\Big].
\end{multline}
Interestingly, the terms in the square brackets vanish because for any $A_i$, 
$\rho^{T_2}_{2S}(A_i)=\rho_{2S}(A_i)$ and $\rho^{T_2}_S(A_i)=\rho_{S}(A_i)$. 
As a consequence, {\it ${\cal E}_{A_1:A_2}$ depends only on the number of singlets 
$n_{A_1:A_2}$ shared between $A_1$ and $A_2$}. Physically, this could have been expected because
the negativity is a measure of the {\it mutual} entanglement between $A_1$ and $A_2$. 
In conclusion, we have
\begin{equation}
\label{gg1}
\mathcal{E}_{A_1 : A_2}={n_{A_1: A_2}}\ln\textrm{Tr}\big|\rho_{2S}^{T_2}
(A_1\cup A_2)
\big|,  
\end{equation}
and using the explicit form of $\rho_{2S}^{T_2}$ \eqref{rhot2}
\begin{equation}
\label{neg-rsp}
\mathcal{E}_{A_1 : A_2}={n_{A_1 : A_2}}\ln 2.
\end{equation}
For a bipartition, i.e., for $A_2 \equiv \bar{A_1}$, with $\bar A_1$ the complement of $A_1$, 
Eq.~\eqref{neg-rsp} is equal to the entanglement entropy of $A_1$ (see Ref.~\onlinecite{RefaelMoore2004}).
This is just a consequence of the fact that in the RS phase the R\'enyi 
entropies $S_{\alpha}$ with varying $\alpha$ are all equal\cite{FagottiCalabreseMoore2011}, while in 
the generic case the negativity for a bipartition is always equal to $S_{1/2}$ \cite{vidal-2002,cct-neg-long}.
This is different from the {\it clean} bipartite system (which is conformal invariant) 
in which the R\'enyi  entropies depend non trivially on the index $\alpha$ in a well-known fashion \cite{cc-04,cc-rev}.

Finally, it is interesting to compare~\eqref{neg-rsp} with the mutual information 
between two intervals ${\mathcal I}_{A_1:A_2}=S_{A_1}+S_{A_2}-S_{A_1\cup A_2}$ 
which can be readily  obtained using $S_X=n_{X:\bar X}\ln2$.
A straightforward calculation  yields 
\begin{equation}
\label{mi-scal}
\mathcal{I}_{A_1: A_2}= (n_{A_1:\bar A_1}+ n_{A_2:\bar A_2} -n_{A_1\cup A_2:B}) \ln 2=
2n_{A_1: A_2}\ln 2.
\end{equation}
Interestingly,~\eqref{mi-scal} coincides, apart from a factor $2$, with ${\cal E}_{A_1:A_2}$ in \eqref{neg-rsp}.

\subsection{The moments of the partially transposed reduced density matrix} 
\label{SECTIONMomentsRSP}

It is instructive to consider the moments $M_\alpha^{T_2}$ of the partially 
transposed density matrix
\begin{equation}
\label{moments}
M_\alpha^{T_2}\equiv\textrm{Tr}\big(\rho_A^{T_2}\big)^\alpha, 
\end{equation}
The logarithmic negativity can be obtained from~\eqref{moments} via the analytic 
continuation $\alpha\to 1$ restricted to the even $\alpha$, as it is routinely done in 
CFT calculations~\cite{calabrese-2012,cct-neg-long}. 
Although $M^{T_2}_{\alpha}$  are not entanglement measures, they encode universal information 
about critical systems. 
In particular, the moments  $M_\alpha^{T_2}$ contain more information than the 
entanglement negativity, and, at least in principle, they could be used to characterize the entire spectrum 
of $\rho_A^{T2}$ on the same lines as it has been done for reduced density matrix~\cite{calabrese-2008}.

The computation of the moments starts by rewriting~\eqref{rhoA} as
\begin{equation}
\rho_A = \Big\{\prod_{k}\bigotimes_{i=1}^{n_{A_k: A_k}}
\rho_{2S}\Big\}
\bigotimes_{i=1}^{n_{A_1: A_2}} \rho_{2S} 
\Big\{\prod_{k}\bigotimes_{i=1}^{n_{A_k :B}}\rho_{S}
\Big\}.
\end{equation} 
Using that the partial transposition acts trivially on the terms in the curly brackets, 
one obtains
\begin{gather}
\label{rhoAT2} 
\rho^{T_2}_A=\Big\{\prod_{k}\bigotimes_{i=1}^{n_{A_k: A_k}}
\rho_{2S}\Big\}\bigotimes_{i=1}^{n_{A_1: A_2}} \rho^{T_2}_{2S} 
\Big\{\prod_{k}\bigotimes_{i=1}^{n_{A_k :B}}\rho_{S}
\Big\}.
\end{gather}
The only two non-zero eigenvalues of $\rho_{A}^{T_2}$ are   
\begin{equation}
\label{eig}
\lambda_{\pm} = \pm  2^{-n_{A: B} - n_{A_1 : A_2}}. 
\end{equation}
The corresponding degeneracies $d_\pm$ are 
\begin{align}
\label{deg}
& d_-= (2^{n_{A_1 : A_2 }} -1) 2^{n_{A_1:A_2 } -1}, \\ 
& d_+= 2^{n_{A: B} + 2 n_{A_1 : A_2}} - d_-. 
\end{align}
The moments $M_\alpha^{T_2}$ can be readily written down as
\begin{equation} 
\label{trrhoN}
M_\alpha^{T_2}= 2^{(n_{A:B} + n_{A_1: A_2}) (1-\alpha)}
\begin{cases}
 2^{n_{A_1: A_2}}  &   \alpha \,\text{even}\\
1                  &   \alpha \, \text{odd}. 
\end{cases}
\end{equation}
In Eq.~\eqref{trrhoN} it is evident that for generic $\alpha$, $M_\alpha^{T_2}$ 
cannot be measures of the entanglement between $A_1$ and $A_2$, since they 
depend on the number of singlets shared with $B$. 
By analytically continuing Eq. \eqref{trrhoN} from the even sequence we recover the 
negativity~\eqref{neg-rsp}, i.e., denoting with $n_e$ the even integers, we have
\begin{equation}
{\cal E}=\lim_{n_e\to 1} \ln M_{n_e}^{T_2},
\end{equation}
while the replica limit of the odd sequence gives the normalization 
${\rm Tr} \rho_A^{T_2}=1$~\cite{calabrese-2012,cct-neg-long}.

\section{Scaling  of the negativity in the random singlet phase}
\label{scal-neg}

In this section we derive the scaling properties of the logarithmic negativity between two 
intervals $A_1$ and $A_2$ in the random singlet phase. 
From the previous section, it is obvious that the only needed ingredient is the 
scaling of the average number of singlets $\langle n_{A_1:A_2}\rangle$ 
shared between $A_1$ and $A_2$ (cf.~\eqref{neg-rsp}). 
Here $\langle\cdot\rangle$ denotes the disorder average. 
Remarkably, $\langle n_{A_1:A_2}\rangle$ can be obtained from the result for a {\it single} interval \cite{RefaelMoore2004}. 
Given an interval $X$ of length $\ell_{X}$ embedded in the infinite line, 
the average number of shared singlets $\langle n_{X:\bar X}\rangle$  scales for large $\ell_X$ as~\cite{RefaelMoore2004} 
\begin{equation}
\label{n_b}
\langle n_{X:\bar X}  \rangle = \frac{1}{3}\ln \ell_{X} + k,
\end{equation}
where $k$ a non-universal constant. 

In order to derive $\langle n_{A_1:A_2}\rangle$, let us consider a generic 
multipartition of the chain into $2k$ blocks as $\cup_{Y\in\Omega_0}Y$, 
with $\Omega_0 = \{ A_1, B_1, \cdots A_k,  B_k \}$ (the case with $k=2$ is in Fig. \ref{figure2} (a)). 
It is convenient to define the set $\Omega$ of all possible compact subintervals of the chain. 
For instance, for $\Omega_0=\{A_1,B_1,A_2,B_2\}$ one has $\Omega=\{A_1,B_1,A_2,B_2,A_1\cup B_1,B_1\cup A_2\}$.  
For each $X \in \Omega$, one can decompose the number of singlets $n_{X:\bar X}$ as the sum 
of all the singlets emerging from an arbitrary block $Y\in X$ and ending in an arbitrary block $Z\in \bar X$, i.e. in formula  
\begin{equation}
\label{decomp}
n_{X: \bar{X}} = \sum_{Y, Z \in \Omega_0} n_{(X \cap Y):(\bar X \cap Z)}.
\end{equation}
After taking the disorder average,~\eqref{decomp} gives 
\begin{equation} 
\label{system}
\frac{1}{3} \ln\ell_{X}+k =\sum_{Y,Z \in \Omega_0} \langle n_{(X \cap Y):(\bar X \cap Z)} \rangle, 
\end{equation}
where we used that~\eqref{n_b} is valid for any $X\in\Omega$. 
One can obtain $\langle n_{X:Y}\rangle$ for any pair $X,Y\in \Omega_0$ by solving the linear system of equations in~\eqref{system} 
generated by considering all $X\in\Omega$, as we are going to show in the following for two adjacent and disjoint intervals.

\subsection{Two adjacent intervals}
\label{sec::adj}

For two adjacent inverals we have $\Omega_0=\Omega= \{A_1, A_2, B_2 \}$, and so the 
system of equations~\eqref{system} becomes 
\begin{equation}
\label{sys}
\begin{cases}
\langle  n_{A_1:A_2} \rangle  + \langle  n_{A_1:B_2}  
\rangle  = \frac{1}{3}\ln (\ell_1),\\
\langle n_{A_2:A_1} \rangle + \langle n_{A_2:B_2} 
\rangle  =\frac{1}{3}\ln (\ell_2),\\
\langle n_{B_2:A_1} \rangle +  \langle n_{B_2:A_2} 
\rangle  =\frac{1}{3}\ln (\ell_1 + \ell_2 ),
\end{cases}
\end{equation}
and the solution for $\langle n_{A_1:A_2}  \rangle$ is 
\begin{equation}
\label{adjacent}
\langle n_{A_1:A_2}  \rangle = \frac{1}{6} 
\ln \left( \frac{\ell_1 \ell_2}{ \ell_1 + \ell_2}\right).
\end{equation}
Consequently [from~\eqref{adjacent} and~\eqref{neg-rsp}], one recovers the scaling  of the logarithmic negativity in 
\eqref{adj-intro}: 
\begin{equation}
\label{neg_adj}
\mathcal{E}_{A_1:A_2}=\frac{\ln 2}{6} 
\ln \Big( \frac{\ell_1 \ell_2}{\ell_1 + \ell_2}\Big)+k,
\end{equation}
%

\subsection{Two disjoint intervals}
\label{sec::disj}

Let us consider the case of two disjoint intervals (see Fig.~\ref{figure2} (a)) 
of lengths $\ell_1,\ell_2$, and at distance $r$, for which $\Omega = \{ A_1, A_2 , B_1, B_2, A_1 \cup B_1 , B_1 \cup A_2\}$. 
From~\eqref{system} one obtains a linear system of six equations with solution for $\langle n_{A_1:A_2}\rangle$ given by
\begin{equation} 
\label{disjoint}
\langle n_{A_1:A_2} \rangle = \frac{1}{6} \ln  \frac{(r + \ell_1)
(r+ \ell_2)}{r (\ell_1 + \ell_2 +r)} .
\end{equation}
Interestingly,~\eqref{disjoint} does not depend on the non-universal additive constant $k$ 
indicating that the logarithmic negativity between two disjoint intervals is a {\it universal scale invariant quantity}, as in the clean case. 
In particular, from~\eqref{disjoint}, ${\cal E}_{A_1:A_2}$ 
can be rewritten as a function only of the cross ratio $x$  as in \eqref{dis-intro}. 

An intriguing consequence of~\eqref{disjoint} is that the 
entanglement between two disjoint intervals decays as $\propto r^{-2}$, i.e.,  with a 
power law of their mutual distance, in stark contrast with the CFT case where this 
decays is exponential~\cite{calabrese-2012}. 
The result for adjacent intervals \eqref{adjacent} is recovered from~\eqref{disjoint} by taking 
$r$ of the order of the lattice spacing (fixed to $1$ in all above formulas) and then considering 
$\ell_{1,2}$ much larger than that.

\section{Numerical SDRG}
\label{n-sdrg}

In this section we present numerical evidence confirming the analytical results 
for the average number of singlets between two arbitrary intervals (adjacent and disjoint)
and consequently for the negativity and mutual information between them. 
In order to do so, we numerical implement the SDRG for finite-size spin chains which works according to the following the steps: 
a) for a given disorder realization, we apply the Dasgupta-Ma rule \eqref{DMrule}, 
i.e., we pair the spins interacting with the strongest bond to form a $SU(2)$ singlet; 
b) the two spins are then decimated, and the remaining couplings renormalized according to~\eqref{ren-coup}; 
c) this SDRG step is iterated until all the spins are paired in singlets. 
At every RG step we monitor the distribution of the renormalized couplings, as well as the spin configurations. 
In this section we only consider a flat probability distribution of the coupling $J$ between 0 and 1.

\subsection{SDRG flow of the renormalized couplings}
\label{rg-flow}

\begin{figure}[t] 
\includegraphics[width=0.9\linewidth]{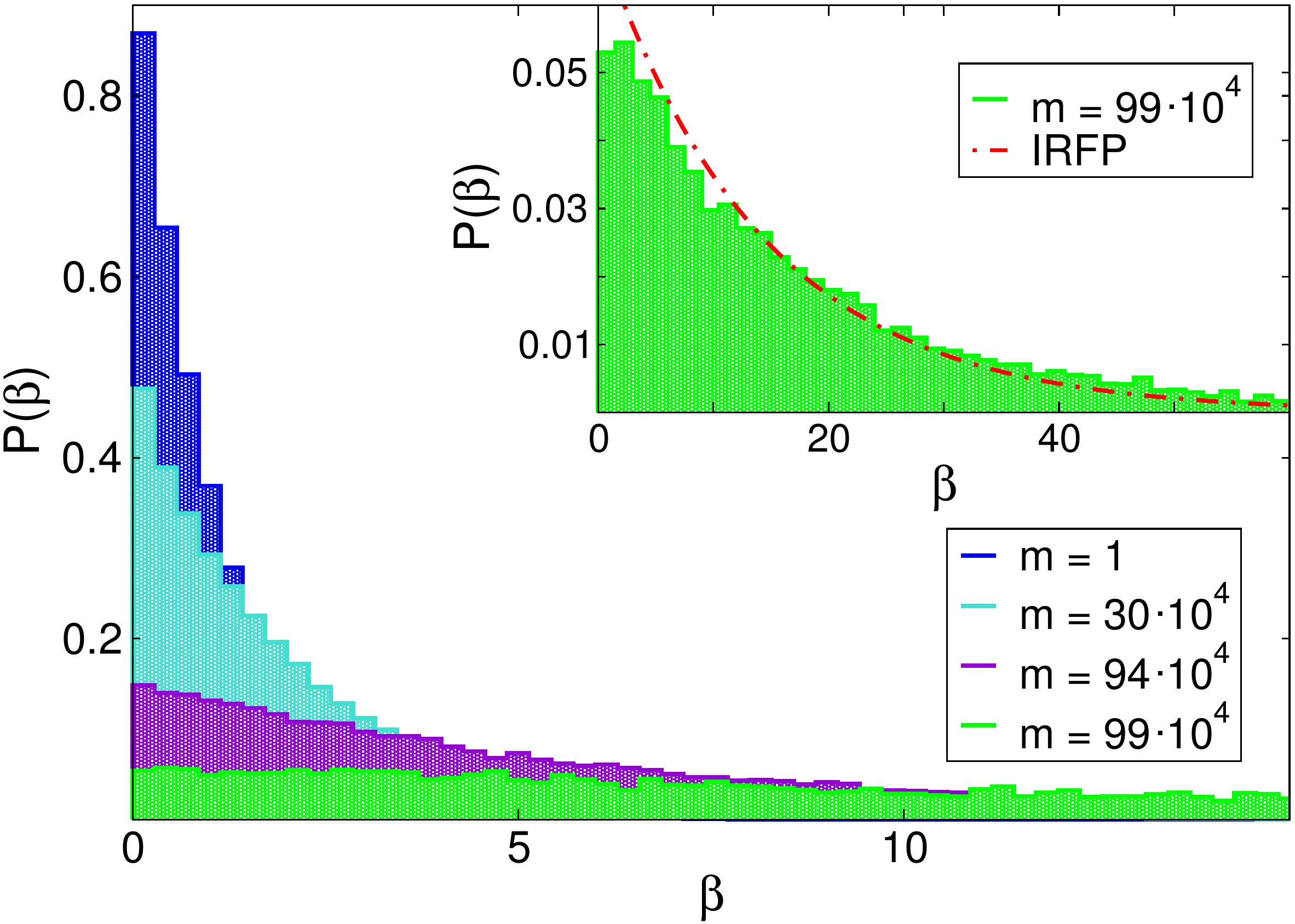}
\caption{ Numerical strong disorder renormalization group: 
 flow of the couplings probability distribution $P(\beta^{
 {}_{(m)}})$ as a function of the SDRG step $m$. Different histograms 
 correspond to different $m$. Results are for the Heisenberg spin 
 chain with $L=10^6$ sites. The width of $P(\beta^{{}_{(m)}}_{i})$ 
 increases upon increasing $m$. Inset: $P(\beta^{{}_{(m)}})$ 
 for $m = 99\cdot 10^4$, compared with the infinite-randomness 
 fixed point (IRFP) result $P^*(\beta) = e^{-\beta/\Gamma}/\Gamma$, 
 (dash-dotted line) with $\Gamma^{{(m)}}$ given in~\eqref{beta_Gamma}.
}
\label{Beta_Flow_final}
 \end{figure}

As a preliminary check of the numerical SDRG, we study the flow of the 
couplings $J_i$ in \eqref{xxz-ham} as a function of the SDRG step $m$. 
The results are reported in Fig.~\ref{Beta_Flow_final}, where we plot the probability distribution $P(\beta_i^{{}_{(m)}})$, 
with  $\beta_i^{{}_{(m)}}\equiv\ln(J_M^{{}_{(m)}}/J^{{}_{(m)}}_i)$ (cf.~\eqref{beta_Gamma}), and $J^{{}_{(m)}}_{M}$ 
the maximum coupling at step $m$. 
The data are for a finite chain with $L=10^6$. 
The initial values of the couplings are drawn from the box distribution $[0,1]$.

Fig.~\ref{Beta_Flow_final} demonstrates the broadening of the couplings 
distribution upon increasing the  SDRG step $m$,  confirming that the SDRG procedure is asymptotically exact at 
large scale~\cite{Fisher1994}. 
The convergence to the universal IRFP distribution $P^*(\beta)$ \eqref{Pfixedpoint} 
is verified in the inset of Fig.~\ref{Beta_Flow_final}.

\begin{figure*}[t]
\includegraphics[width=0.9\linewidth]{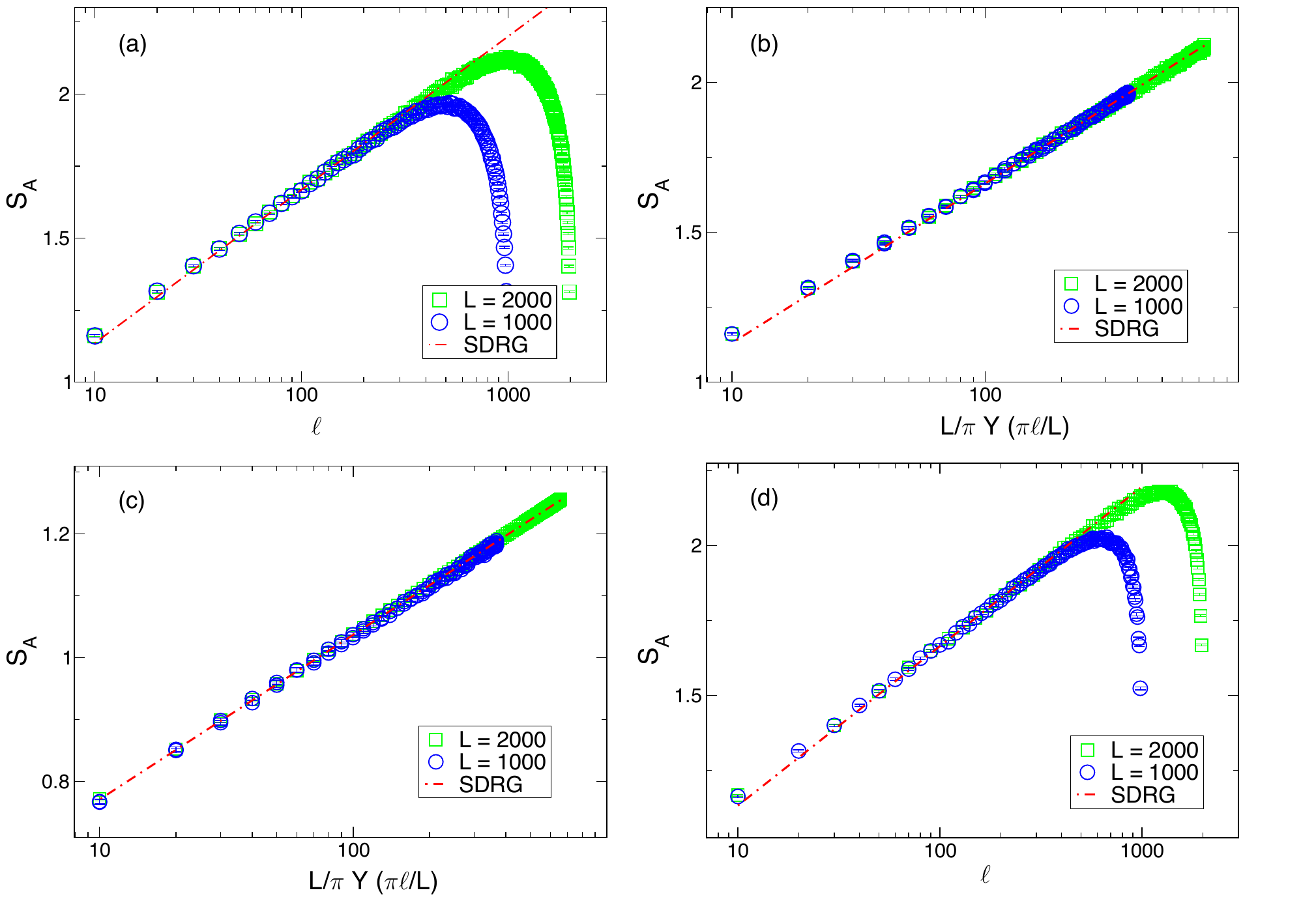}
\caption{The single interval entanglement entropy for the disordered Heisenberg spin chain.
 The data correspond to the average over $73000$  disorder realizations. 
 Notice that (here and in the following figures) we report also the statistical errors, 
 although the error bars are smaller than the symbol sizes. 
 Different symbols correspond to different chain sizes $L$ and dashed dotted lines are one-parameter fit to 
 the analytic prediction from the SDRG. 
 {(a)} A single interval in a periodic chain: $S_A$ plotted versus the interval length $\ell$. 
 {(b)} Same data as in (a) plotted against the modified chord length \eqref{chord-l}.
 {(c)} A single interval starting from the boundary of a chain with OBC versus the modified chord length \eqref{chord-l}.
 {(d)} A single interval in the bulk of an open chain. 
}
\label{fig_SI}
\end{figure*}

\subsection{Single-interval entanglement entropy}
\label{1block-ent}

Another important check of the validity of the numerical SDRG is provided 
by the scaling of the single block entanglement entropy. 
We consider the random $XXX$ chain with both open and periodic boundary conditions. 
For OBC we expect different scalings depending on whether the block $A$ is in the bulk of the chain 
or whether it touches the boundary.
The entanglement entropy $S_A$ is obtained by determining the number of singlets between $A$ and $B$ for each realization 
of the disorder and then by multiplying the average by $\ln2$. 
We  report the variance of the distribution as an estimate of statistical error.  

Numerical SDRG results for $S_A$ are shown in Fig.~\ref{fig_SI} (a)-(d). 
In all panels the data are for chains of length $L=1000$ and $L=2000$. The 
disorder average is over $73000$ disorder realizations.
We start by considering in panel (a)  the block $A$ in a periodic chain. 
For $\ell\lesssim L/2$, $S_A$ increases logarithmically as  function of the interval length $\ell$ (note 
the logarithmic scale on the $x$-axis). 
Instead, for $\ell\gtrsim L/2$ finite-size effects are visible. 
The dash-dotted line in the figure is a fit to the expected SDRG result~\cite{RefaelMoore2004}
\begin{equation}
\label{fit}
S_A=\frac{\ln2}{3}\ln\ell +K, 
\end{equation}
with $K$ the only  fitting parameter. For $\ell\lesssim L/2$ the data are in perfect agreement with~\eqref{fit}. 

As discussed in Ref.~\onlinecite{FagottiCalabreseMoore2011},
finite-size effects can be taken into account by replacing 
in~\eqref{sa}  the interval length $\ell$ with the \emph{modified} chord length  $L_c$ as 
\begin{equation}
\label{chord-l}
\ell \to L_c\equiv\frac{L}{\pi} Y \Big(\frac{\pi\ell}{L}\Big).
\end{equation}
Here $Y(x)$ is a symmetric function under $x\to\pi-x$, which ensures 
$S(\ell)=S(L-\ell)$. It has been found that for a single block in a periodic chain, 
$Y(x)$ is well approximated by \cite{FagottiCalabreseMoore2011}
\begin{equation}
Y(x)=\sin(x)\Big(1+\frac{4}{3} k_1 \sin^2 (x)\Big), 
\label{Yx}
\end{equation}
with $k_1 \sim 0.115$. Using~\eqref{chord-l} in~\eqref{fit} one obtains 
\begin{equation}
\label{fit1}
S_A=\frac{\ln2}{3}\ln(L_c)+K. 
\end{equation}
Fig.~\ref{fig_SI} (b) reports the data for $S_A$ versus $L_c$. The dash-dotted line in Fig.~\ref{fig_SI} (b) 
is a fit to~\eqref{fit1}, and it is in perfect agreement with the SDRG data for all values of $\ell$ confirming 
the correctness of~\eqref{fit1}.

In Fig.~\ref{fig_SI} (c) we consider an open chain with the block $A$ starting from the boundary. 
For a semi-infinite system, the entanglement entropy is expected to be~\cite{RefaelMoore2004} 
\begin{equation}
\label{sa}
S_A=\frac{\ln2}{6}\ln(\ell) +K', 
\end{equation}
where the prefactor of the logarithm takes into account that subsystem $A$ shares only one edge with its complement. 
This formula indeed describes accurately the data for $\ell\lesssim L/2$.
The finite size and boundary effects can be taken into account again by replacing $\ell$ by a modified chord length \eqref{chord-l}. 
We find that the same function $Y(x)$ in Eq. \eqref{Yx} describes very accurately 
the data for OBC as shown in Fig.~\ref{fig_SI} (c). 
This is a non trivial result since there is no conformal invariance to guarantee the equality of the two
finite size scaling functions (i.e. open and periodic) as in the clean case. 

Finally in Fig.~\ref{fig_SI} (d) we consider a block of length $\ell$ in the center of an open chain. 
For $\ell\lesssim L/2$ we expect that the block $A$ does not feel significantly the presence of the boundary 
and \eqref{fit} should describe accurately the data, as evident from the figure.
It is instead not obvious how to modify the prediction to take into account finite-size and boundary effects:
we only mention that replacing $\ell$ with the modified chord length \eqref{Yx} does not work. 
This does not come unexpected since also in clean systems the results is more complicated~\cite{x11}. 

\begin{figure}[t] 
\includegraphics[width=0.85\linewidth]{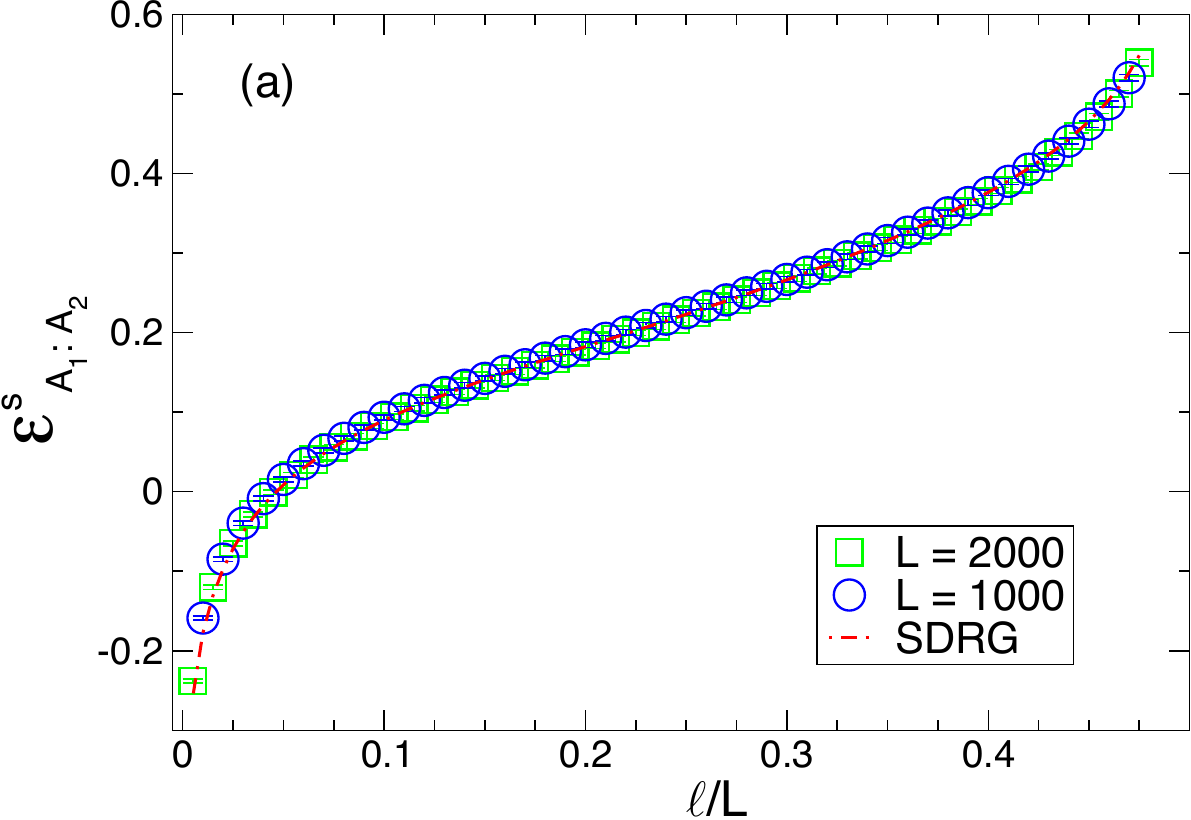}
\includegraphics[width=0.85\linewidth]{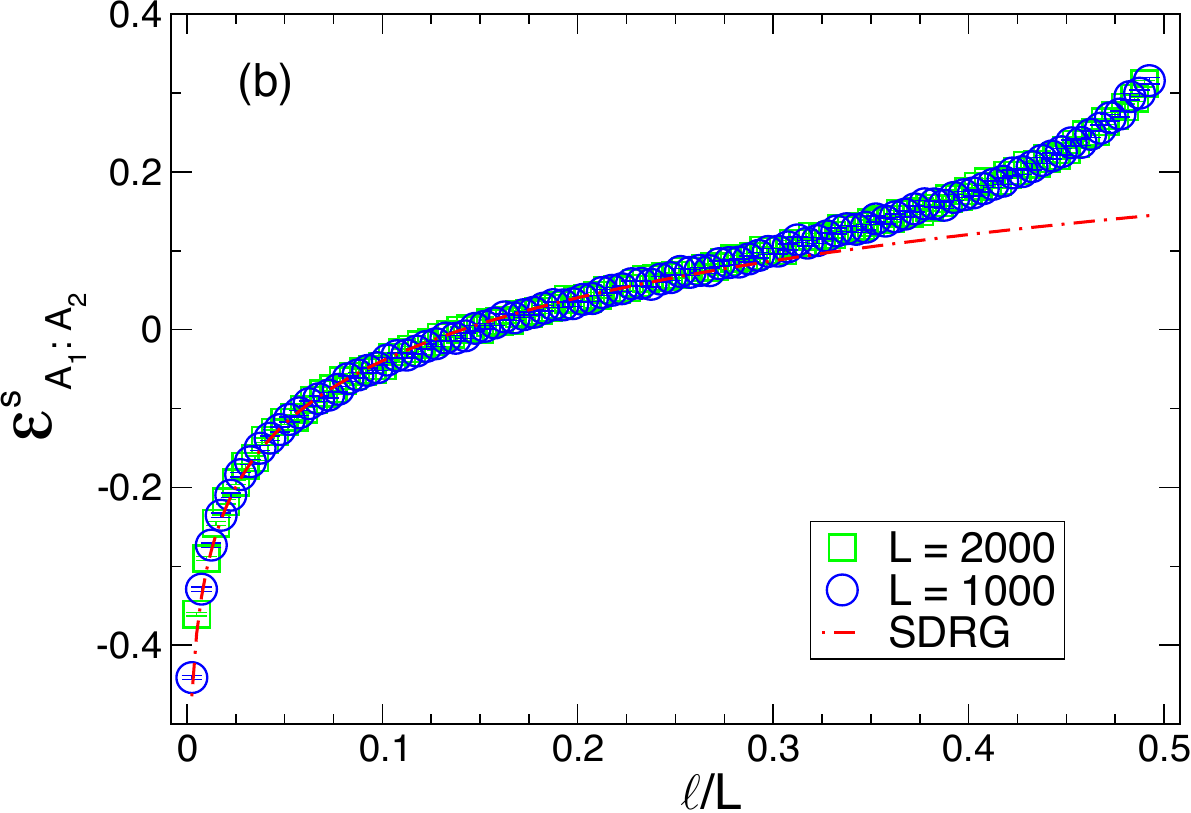}
\caption{Shifted logarithmic negativity $\mathcal{E}^s_{A_1 : A_2}$ \eqref{Es}
between two adjacent intervals of equal-length $\ell$ in the random Heisenberg  chain. 
 ${\cal E}^s_{A_1:A_2}$ is plotted against $\ell/L$. 
 The symbols correspond to the average over $73000$ disorder realizations. 
 Different symbols correspond to different system sizes. 
 (\textcolor{red}{a}): For a periodic chain the data are perfectly described by \eqref{AIY} with only 
 one fitting parameter.   
 (\textcolor{red}{b}): For an open chain, we can only use the prediction for the infinite chain which describes 
 well the data as long as $\ell\ll L$. 
}
\label{fig_2AI}
\end{figure}

\subsection{Logarithmic negativity: Two adjacent intervals}
\label{rsp-negativity}

Using the numerical SDRG method, we now study the 
scaling behavior of the logarithmic negativity ${\cal E}_{A_1:A_2}$ between 
two adjacent intervals $A_1$ and $A_2$ in the random Heisenberg chain. 
Specifically, we provide robust numerical evidence for~\eqref{neg_adj}. 
The negativity is just obtained from the statistics of the singlets between 
$A_1$ and $A_2$ and multiplying the resulting average by $\ln2$.

In Fig.~\ref{fig_2AI} we report the SDRG data for ${\cal E}_{A_1:A_2}$ for two adjacent intervals of the same length $\ell$. 
Fig.~\ref{fig_2AI} (a) shows the SDRG results for a periodic
chain, while Fig.~\ref{fig_2AI} (b) is for  two intervals in the middle of an open chain. 
We plot the shifted negativity 
\begin{equation}
{\cal E}^s_{A_1:A_2}= {\cal E}_{A_1:A_2}-\frac{\ln2}6 \ln L,
\label{Es}
\end{equation}
as a function of $\ell/L$ so that data for different chain lengths are expected to collapse on a single scaling curve.  
This clearly happens  for both open and periodic chains. 
In both cases ${\cal E}_{A_1:A_2}$ increases logarithmically for $\ell\lesssim L/4$, when finite-size effects kick in. 
Actually, for PBC one can take into account all the finite-size effects by replacing 
all lengths in \eqref{adj-intro} with modified chord lengths, obtaining 
\begin{equation}
{\cal E}^s_{A_1:A_2}\simeq \frac{\ln 2}{6}\ln\frac{Y_c^2(\pi\ell/L)}{Y_c(2\pi\ell/L)}+k.
\label{AIY}
\end{equation}
The SDRG are perfectly described by this prediction as clear from Fig. \ref{fig_2AI} (a) where 
the dashed line is a one parameter fit to \eqref{AIY}.

Conversely, for the OBC chain in Fig.~\ref{fig_2AI} (b) there is not a simple modification of the result for infinite chain 
to take into account the boundary effects, as it was the case also for clean systems~\cite{ctc-14}.
For this reason, in the figure we only report the one-parameter fit to the SDRG prediction for the infinite chain~\eqref{neg_adj}
valid for $\ell\ll L$.

\subsection{Logarithmic negativity: Two disjoint intervals}

\begin{figure}[t] 
\includegraphics[width=0.85\linewidth]{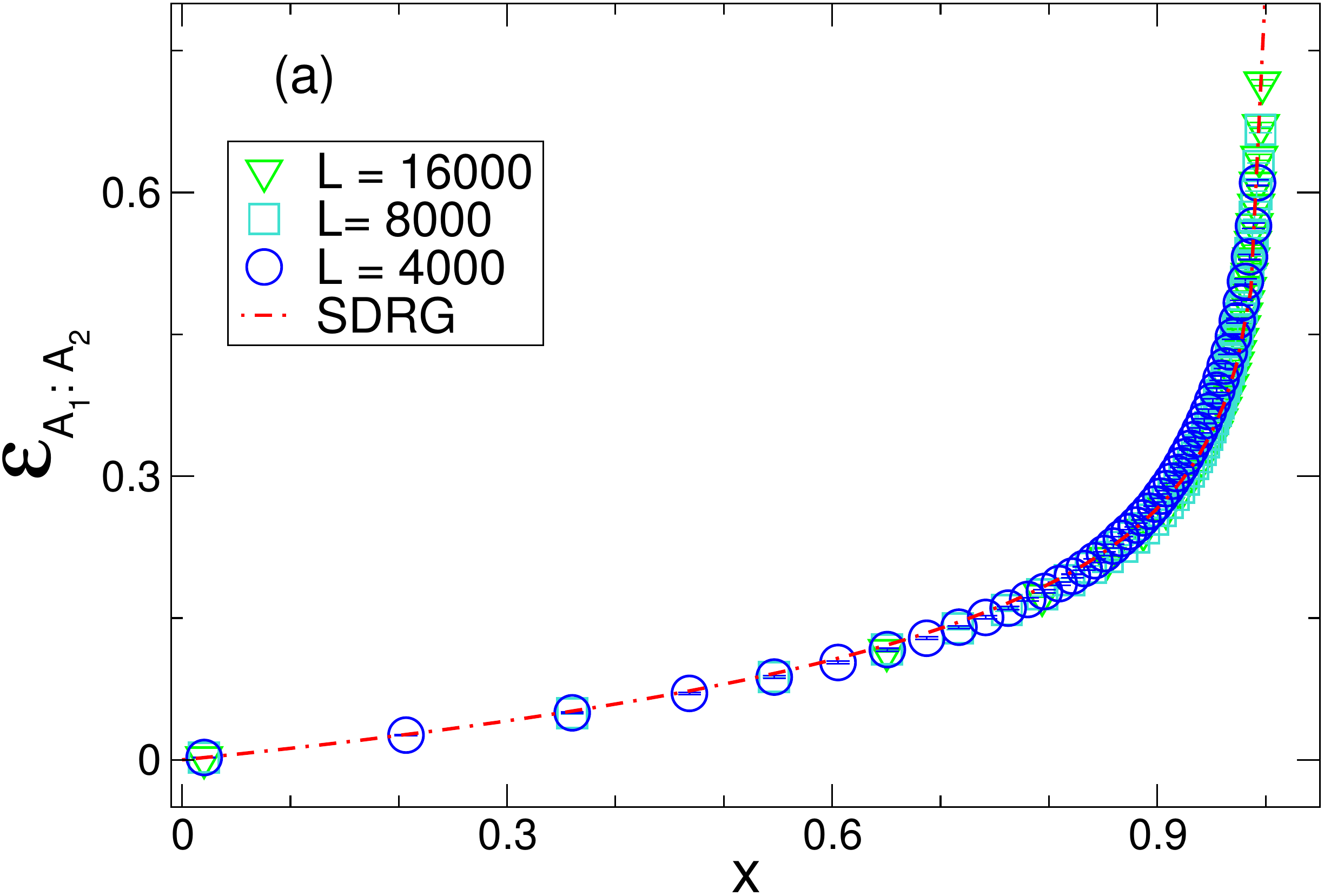}
\includegraphics[width=0.85\linewidth]{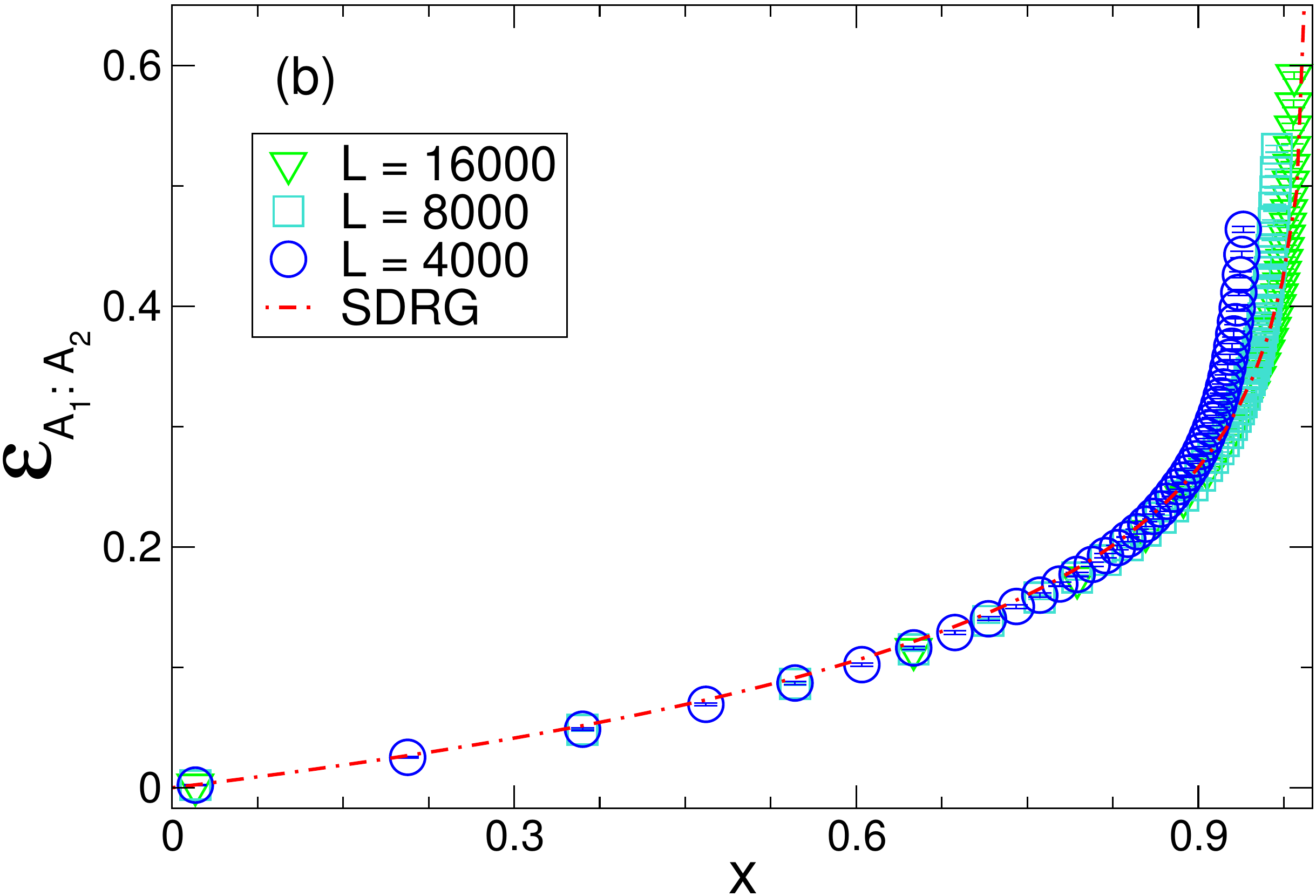}
\caption{Logarithmic negativity $\mathcal{E}_{A_1:A_2}$ between two disjoint intervals of equal length $\ell$ 
 in the random Heisenberg spin chain plotted against the cross ratio $x$.
 The data are averaged over $73000$ disorder realizations. 
 Different symbols correspond to different system sizes. 
 (\textcolor{red}{a}): For a periodic chain the cross-ratio $x$ is given by \eqref{Ycc} and  the dashed line is the 
 analytic SDRG prediction \eqref{dis1}. 
 (\textcolor{red}{b}): For an open chain  we consider the infinite volume definition of the cross ratio \eqref{xdef} 
 and  the dash-dotted line is the SDRG prediction \eqref{dis1} to which the data tend for large chains. 
}
\label{fig_2DI}
\end{figure}

We finally move to the most interesting case of the logarithmic negativity between two  disjoint intervals of equal length
$\ell$ at distance $r$.
As usual in this section, we compute the negativity from the statistics of the singlets between the two intervals. 
Our results are reported in Fig.~\ref{fig_2DI} for both periodic and open chains.  

In the case of a periodic chain, we plot $\mathcal{E}_{A_1:A_2}$ as a function of 
the cross-ratio $x$ \eqref{xdef} in which we have substituted all lengths by the corresponding modified chord lengths, i.e.
\begin{equation}
x=\frac{Y^2_c(\pi\ell/L)}{Y^2_c(\pi(\ell+r)/L)},
\label{Ycc}
\end{equation}
in order to take into account finite-size effects. 
The SDRG asymptotic prediction for the negativity is given by Eq. \eqref{dis-intro} which reads 
\begin{equation}
{\cal E}_{A_1:A_2}=-\frac{\ln2}{6}\ln (1-x). 
\label{dis1}
\end{equation}
In Fig. \ref{fig_2DI} (a) the data correspond to a fixed choice of $r$ and $\ell$ is running up to $(L-r)/2$.
It is clear from the figure that the prediction (\ref{dis1}) describes incredibly well the numerical data, 
without any fitting parameter. 
Notice that $x$ close to $0$ corresponds to far away intervals and the decay of the negativity is algebraic,  
in constrast with the exponential behavior of clean systems\cite{calabrese-2012,cct-neg-long}.
Oppositely, the limit $x\to1$ corresponds to very close intervals and the logarithmic divergence is needed to reproduce
the adjacent intervals results.

For the case of an open chain, we limit ourselves to consider two intervals centered around the middle of the chain, i.e. 
$A_1=[-\ell-r/2,-r/2]$ and $A_2=[\ell+r/2,r/2]$.
We generate the data by fixing a value of $r$ and letting $\ell$ run up to the boundary. 
In this case, we do not have a prediction which takes into account finite size effects, 
so we expect Eq.~\eqref{dis1} to describe correctly the data for $\ell,r\ll L$, i.e. when the 
effects of the boundary can be neglected. 
For this reason in Fig.  \ref{fig_2DI} (b) we report the negativity as function of the cross-ratio $x$
given by the infinite volume formula $x=(\ell/(\ell+r))^2$. 
It is evident from the figure that this prediction describes correctly the data for $x$ far from $1$, as it should. 
With increasing $L$, the effects of the boundaries becomes less and less relevant and the 
data get closer to the infinite volume result, as expected.

\section{Logarithmic negativity in DMRG simulations}
\label{neg-dmrg}

\begin{figure}[t]
\includegraphics*[width=0.99\linewidth]{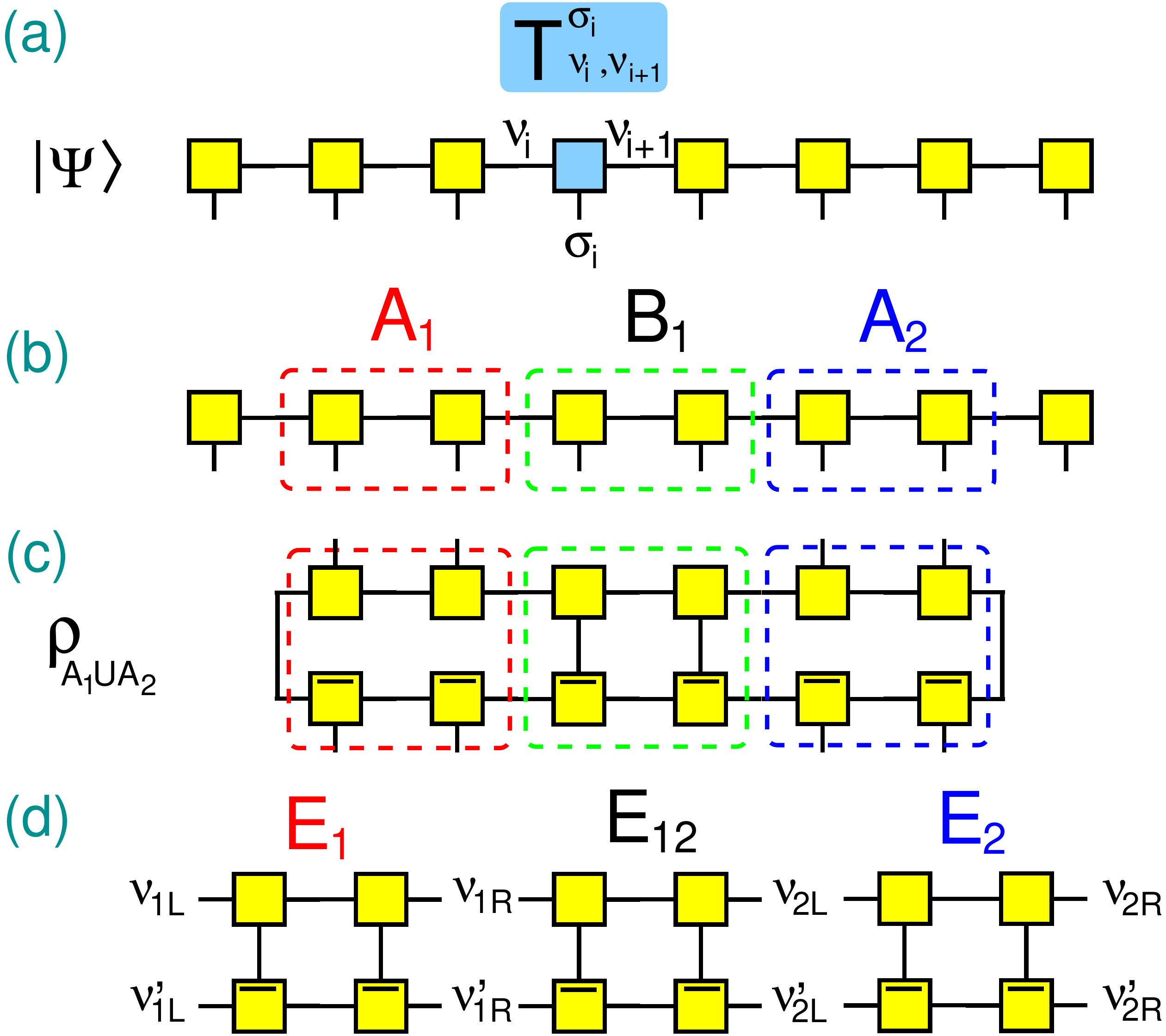}
\caption{ Reduced density matrix with Matrix Product States (MPS) for 
 a chain with $L=8$ sites and open boundary conditions. (a) The MPS 
 representation for a generic wavefunction $|\Psi\rangle$. The 
 basic tensor $T^{\sigma_i}_{\nu_{i},\nu_{i+1}}$ at site $i$ of the 
 chain is shown in the shaded box. The vertical leg denotes the physical 
 index $\sigma_i$, while the horizontal ones are the virtual indices 
 $\nu_i$ (summed over). Tensors in the bulk have rank $3$, while at 
 the edges have rank $2$. (b) The tripartition of the chain as $A_1\cup B_1\cup A_2\cup B_2$. 
 $A_1$ and $A_2$ are two equal blocks formed by two spins. The two blocks are at distance 
 two (spins in $B_1$). (c) The reduced density matrix $\rho_{A_1\cup A_2}$. 
 Only the tensors living in $A_1\cup B_1\cup A_2$ appear. (d) Definition 
 of the transfer matrices $E_1$, $E_2$, $E_{12}$ corresponding to parts 
 $A_1$, $A_2$, and $B_1$ of the chain. 
}
\label{cartoon1}
\end{figure}

In this section we describe how to calculate the logarithmic negativity in DMRG simulations. 
We consider a generic $1D$ lattice of $L$ sites, restricting ourselves to 
open boundary conditions. On each site $i$ of the lattice we consider a local 
Hilbert space of finite dimension $d$ (e.g., for spin-$1/2$ systems one has 
$d=2$). A generic wavefunction $|\Psi\rangle$ of a pure state is written as 
a Matrix-Product-State (MPS) as 
\begin{equation}
|\Psi\rangle=\sum\limits_{\sigma_1,\dots,\sigma_L}T^{\sigma_1}_{\nu_1}
T^{\sigma_2}_{\nu_1\nu_2}\cdots T^{\sigma_L}_{\nu_L}|\sigma_1,\dots,
\sigma_L\rangle.
\label{mps} 
\end{equation}
Here $\sigma_i$ with $1\le i\le d$ is the so-called physical index representing the 
states in the local Hilbert space, whereas $1\le\nu_i\le\chi_i$ are the virtual 
indices, with $\chi_i$ the local bond dimension of the MPS. In~\eqref{mps} the sum 
over the repeated indices $\nu_i$ is assumed, and one has $|\sigma_1,\sigma_2,\dots,
\sigma_L\rangle\equiv|\sigma_1\rangle\otimes|\sigma_2\rangle\cdots|\sigma_L
\rangle$, with $|\sigma_i\rangle$ an element of the basis of the Hilbert space 
at site $i$. At fixed $\sigma_i$, $T^{\sigma_i}_{\nu_{i}\nu_{i+1}}$ is a 
$\chi_i\times\chi_i$ matrix, while $T^{\sigma_1}_{\nu_1}$ and $T^{\sigma_L}_{\nu_L}$ 
are vectors. The MPS in~\eqref{mps} can be pictorially represented as in Fig.~\ref{cartoon1} 
(a), where the boxes denote the tensors $T_{\nu_i,\nu_{i+1}}^{\sigma_i}$, and horizontal 
and vertical legs the virtual and physical indices, respectively. The full system 
density matrix $\rho\equiv|\Psi\rangle\langle\Psi|$ is obtained as
\begin{multline}
\label{rho}
\rho=\sum\limits_{\substack{\sigma_1,\dots,\sigma_L\\\sigma_1',\dots,\sigma_L'}}
(T^{\sigma_1}_{\alpha_1}\bar T^{\sigma_1'}_{\nu_1'})(T^{\sigma_2}_{\nu_1,\nu_2}
\bar T^{\sigma_2'}_{\nu_1',\nu_2'})\cdots\\\cdots (T^{\sigma_L}_{\nu_L}\bar T^{
\sigma_L'}_{\nu_L'})|\sigma_1\rangle\langle\sigma_1'|\otimes\cdots\otimes|\sigma_L
\rangle\langle\sigma_L'|, 
\end{multline}
where the bar in $\bar T^{\sigma_i}_{\nu_i,\nu_{i+1}}$ (and correspondingly in 
the boxes in Fig.~\ref{cartoon1}) denotes the complex conjugation.

\begin{figure}[t]
\includegraphics*[width=0.99\linewidth]{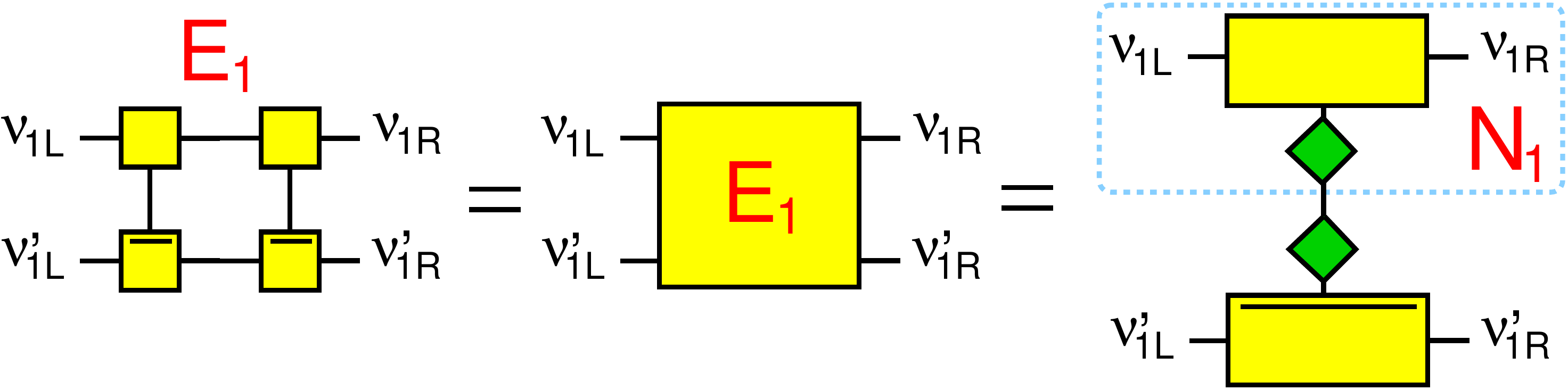}
\caption{ The definition of the tensor $N_1$ (cf.~\eqref{rho-fin}) from the singular 
 value decomposition (SVD) of the transfer matrix $E_1$. The SVD is performed 
 with respect to the index pairs $(\nu_{1L},\nu_{1R})$ and $(\nu'_{1L},\nu'_{1R})$. 
 The rhombi denote the diagonal matrix containing the square root of the singular 
 values. The same definition holds for $N_2$. 
}
\label{cartoon2}
\end{figure}

\subsection{The partially transposed reduced density matrix}
\label{sec::rdm}

In order to calculate the logarithmic negativity we now introduce a tripartition 
of the system (Fig.~\ref{cartoon1}) (b) as $A_1\cup B_1\cup A_2\cup B_2$, where 
$A_1$ and $A_2$ are the intervals of interest and $B_1,B_2$ the remaining parts. 
For simplicity we consider the case of two equal-length intervals with $\ell_1=\ell_2
=\ell$ at mutual distance $r$, and shifted from the left boundary by $s$ sites, i.e., 
$A_1=[s+1,\dots,s+\ell]$, $A_2=[s+\ell+r+1,\dots,s+r+2\ell]$. This is illustrated in 
Fig.~\ref{cartoon1} (b) for $s=1$, $\ell=2$, and $d=2$. The reduced density matrix 
$\rho_{A_1\cup A_2}$ is obtained by tracing over the degrees of freedom of $B_1\cup B_2$ 
in~\eqref{rho} as 
\begin{multline}
\label{rho12}
\rho_{A_1\cup A_2}=\delta_{\nu_{1L},\nu'_{1L}}
\delta_{\nu_{2R},\nu'_{2R}}\\
\times\sum\limits_{\substack{\{\sigma_i,\sigma'_{i}:\, i\in A_1\}}}
[F_1]^{\{\sigma_i\}}_{\nu_{1L},\nu_{1R}}
[\bar F_1]_{\nu'_{1L},\nu'_{1R}}^{\{\sigma'_i\}}\\
\times\sum\limits_{\substack{\{\sigma_i,\sigma'_{i}:\, i\in A_2\}}}
[F_2]^{\{\sigma_i\}}_{\nu_{2L},\nu_{2R}}
[\bar F_2]_{\nu'_{2L},\nu'_{2R}}^{\{\sigma'_i\}}\\
\times[E_{12}]^{\nu_{1R},\nu_{2L}}_{\nu'_{1R},\nu'_{2L}}
\prod\limits_{i\in A_1\cup A_2}
|\sigma_i\rangle\langle\sigma'_i|,
\end{multline}
where we defined 
\begin{equation}
[F_1]^{\{\sigma_i\}}_{\nu_{1L},\nu_{1R}}\equiv\prod\limits_{i
\in A_1}T^{\sigma_i}_{\nu_{i},\nu_{i+1}}, 
\end{equation}
with $\nu_{1L}\equiv\nu_{s+1}$, $\nu_{1R}\equiv\nu_{s+\ell+1}$ (cf. 
Fig.~\ref{cartoon1} (d) for the definition of the indices). A similar definition holds 
for $[F_{2}]^{\{\sigma_i\}}_{\nu_{2L},\nu_{2R}}$. The term $\delta_{\nu_{1L},\nu'_{1L}}
\delta_{\nu_{2R},\nu'_{2R}}$ in~\eqref{rho12} is because the virtual indices $\nu_L,\nu'_L$ 
and $\nu_{1R},\nu'_{1R}$ are contracted. In~\eqref{rho12} $E_{12}$ is the ``transfer matrix'' 
of the region connecting the two intervals ($B_1$ in Fig.~\ref{cartoon1} (c)). 
The definition of $E_{12}$ and of the transfer matrices $E_1$, $E_2$, which are associated 
with the two intervals $A_1$ and $A_2$, are given as 
\begin{align}
\label{E1}
& [E_1]^{\nu_{1L},\nu_{1R}}_{\nu'_{1L},\nu'_{1R}}\equiv
\prod\limits_{j\in A_1}\sum_{\sigma_j}T^{\sigma_j}_{\nu_j,\nu_{j+1}}
\bar T^{\sigma_j}_{\nu'_j,\nu'_{j+1}}\\
\label{E2}
& [E_2]^{\nu_{2L},\nu_{2R}}_{\nu'_{2L},\nu'_{2R}}\equiv
\prod\limits_{j\in A_2}\sum_{\sigma_j}T^{\sigma_j}_{\nu_j,\nu_{j+1}}
\bar T^{\sigma_j}_{\nu'_j,\nu'_{j+1}}\\
\label{E}
& [E_{12}]^{\nu_{1R},\nu_{2L}}_{\nu'_{1R},\nu'_{2L}}\equiv 
\prod\limits_{j\in B_1}\sum_{\sigma_j}T^{\sigma_j}_{\nu_j,\nu_{j+1}}
\bar T^{\sigma_j}_{\nu'_j,\nu'_{j+1}}. 
\end{align}
Notice that all the spin indices are contracted in~\eqref{E1}\eqref{E2}\eqref{E}. 
These definitions are illustrated in Fig.~\ref{cartoon1} (d).

The computational cost to construct $E_1,E_2,E_{12}$ is ${\mathcal O}(\chi^6_{max})$, 
with $\chi_{max}\equiv\textrm{max}_i(\chi_i)$ the largest bond dimension of the MPS. 
It is convenient to introduce a basis for the Hilbert space of the two intervals $A_1$ 
and $A_2$ as 
\begin{align}
|w^{(1)}_{\nu_{1L},\nu_{1R}}\rangle\equiv\sum\limits_{\sigma_i:\,i\in A_1}
[F_1]^{\{\sigma_i\}}_{\nu_{1L},\nu_{1R}}\prod\limits_{j\in A_1}|\sigma_j\rangle,\\
|w^{(2)}_{\nu_{2L},\nu_{2R}}\rangle\equiv\sum\limits_{\sigma_i:\,i\in A_2}
[F_2]^{\{\sigma_i\}}_{\nu_{2L},\nu_{2R}}\prod\limits_{j\in A_2}|\sigma_j\rangle, 
\end{align}
where the index pairs $(\nu_{1L},\nu_{1R})$ and $(\nu_{2L},\nu_{2R})$ label the 
different states in the two bases. This allows one to rewrite~\eqref{rho12} in the 
compact form  
%
\begin{multline}
\label{rho12-r}
\rho_{A_1\cup A_2}=\delta_{\nu_{1L},\nu'_{1L}}\delta_{\nu_{2R},\nu'_{2R}}
[E_{12}]^{\nu_{1R},\nu_{2L}}_{\nu'_{1R},\nu'_{2L}}\\|w^{(1)}_{\nu_{1L},\nu_{1R}}
\rangle\langle w^{(1)}_{\nu'_{1L},\nu'_{1R}}|\otimes |w^{(2)}_{\nu_{2L},
\nu_{2R}}\rangle\langle w^{(2)}_{\nu'_{2L},\nu'_{2R}}|.
\end{multline}
%
Notice that now $\rho_{A_1\cup A_2}$ is at most a $\chi^2_{max}\times\chi^2_{max}$ matrix. 
Importantly, for a generic tripartition the vectors $|w^{{}_{(1)}}_{\nu_{1L},\nu_{1R}} 
\rangle$ and $|w^{{}_{(2)}}_{\nu_{2L},\nu_{2R}}\rangle$ are not orthogonal. Clearly, 
one has the overlap matrices 
\begin{align}
\langle w^{(1)}_{\nu_{1L},\nu_{1R}}|w^{(1)}_{\nu'_{1L},\nu'_{1R}}\rangle&=
[E_1]^{\nu_{1L},\nu_{1R}}_{\nu'_{1L},\nu'_{1R}} \\
\langle w^{(2)}_{\nu_{2L},\nu_{2R}}|w^{(2)}_{\nu'_{2L},\nu'_{2R}}\rangle&=
[E_2]^{\nu_{2L},\nu_{2R}}_{\nu'_{2L},\nu'_{2R}}, 
\end{align}
where $E_1$ and $E_2$ are defined in~\eqref{E1} and~\eqref{E2}, respectively. 

\begin{figure}[t]
\includegraphics*[width=0.99\linewidth]{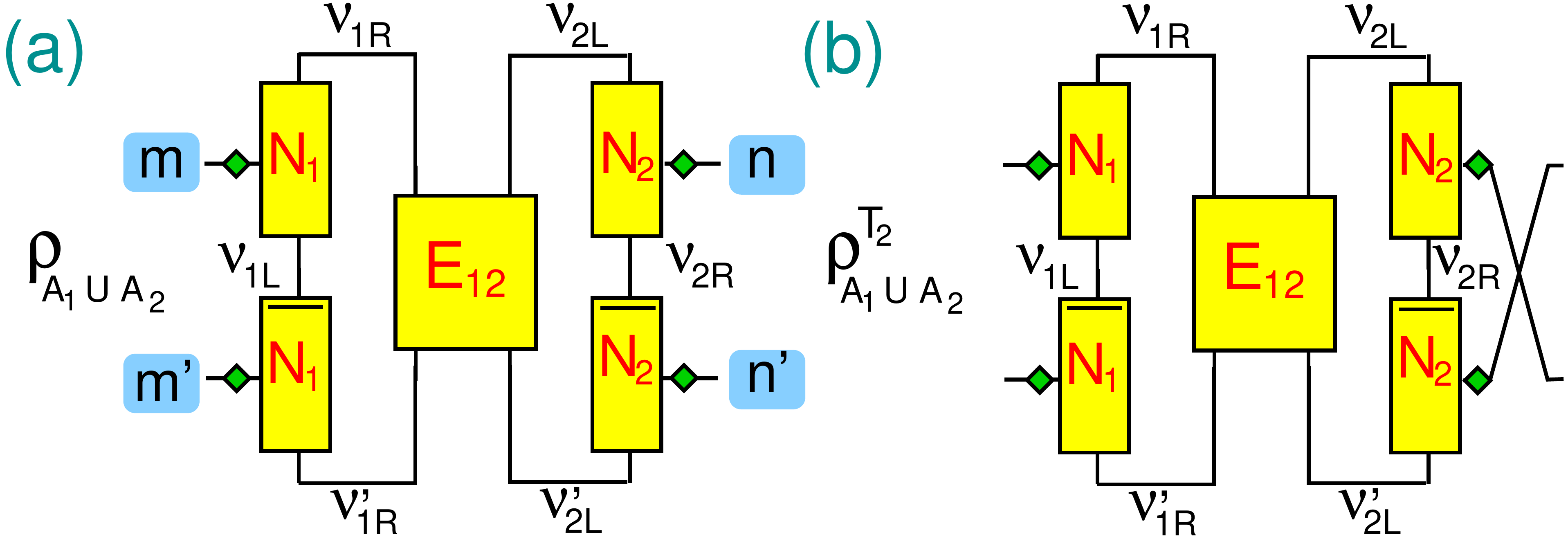}
\caption{ (a) The reduced density matrix of two intervals $\rho_{A_1\cup A_2}$ 
 in the MPS framework. The tensor $N_1$ is defined in Fig.~\ref{cartoon2} (a similar 
 definition holds for $N_2$), whereas $E_{12}$ is as in Fig.~\ref{cartoon1} (d). The 
 indices are as in~\eqref{rho-fin}. The index pairs $(m,m')$ and $(n,n')$ refer to interval 
 $A_1$ and $A_2$, respectively. (b) The partially transposed reduced density matrix $\rho_{A_1
 \cup A_2}^{T_2}$. Here the partial transposition is performed with respect to subsystem 
 $A_2$, and it corresponds to the exchange $n\leftrightarrow n'$.
}
\label{cartoon3}
\end{figure}

Interestingly, when the the two intervals $A_1$ and $A_2$ are close to the edges of the 
chain (this is the geometry considered in Ref.~\onlinecite{hannu-2008}) one has that 
$|w^{{}_{(1)}}_{\nu_{1L},\nu_{1R}}\rangle\to|w^{{}_{(1)}}_{\nu_{1R}}\rangle$ and 
$|w^{{}_{(2)}}_{\nu_{2L},\nu_{2R}}\rangle\to|w^{{}_{(2)}}_{\nu_{2L}}\rangle$, 
with $|w^{{}_{(1)}}_{\nu_{1L}}\rangle$ and $|w^{{}_{(2)}}_{\nu_{2L}}\rangle$ forming 
two orthonormal bases. Thus, it is straigthforward to show that $\rho_{A_1\cup A_2}$ is 
simply given as 
\begin{equation}
\label{han}
\rho_{A_1\cup A_2}=[E_{12}]^{\nu_{1R},\nu_{2L}}_{\nu'_{1R},\nu'_{2L}} \,
|w^{{}_{(1)}}_{\nu_{1R}}\rangle\langle w^{{}_{(1)}}_{\nu'_{1R}}|\otimes 
|w^{{}_{(2)}}_{\nu_{2L}}\rangle\langle w^{{}_{(2)}}_{\nu'_{2L}}|. 
\end{equation}
The spectrum of $\rho_{A_1\cup A_2}$ is obtained by diagonalizing $E_{12}$, 
with computational cost $\propto\chi_{max}^6$. 
Moreover, $\rho^{T_2}_{A_1\cup A_2}$ is obtained from~\eqref{han} by 
exchanging $\nu_{2L}\leftrightarrow\nu'_{2L}$.

For generic tripartitions it is convenient to orthonormalize 
the vectors $|w^{{}_{(1)}}_{\nu_{1L},\nu_{1R}}\rangle$ and $|w^{{}_{(2)}}_{
\nu_{2L},\nu_{2R}}\rangle$. This can be done via a singular value decomposition 
(SVD) of the transfer matrices $E_1$ and $E_2$ (cf.~\eqref{E1}\eqref{E2}). Precisely, 
given the SVD of $E_1$ as $E_1=UDV^\dagger$ with $U,V$ unitaries and $D$ diagonal, the 
vectors $|v^{{}_{(1)}}_m\rangle$ defined as 
\begin{equation}
\label{v1}
|v^{{}_{(1)}}_{m}\rangle\equiv [\widetilde N_1]^m_{\nu_{1L},\nu_{1R}}
|w^{{}_{(1)}}_{\nu_{1L},\nu_{1R}}\rangle, 
\end{equation}
with $\widetilde N_1\equiv D^{-\frac{1}{2}}U^\dagger$, are orthonormal. 
Similarly, for $A_2$ one has the orthonormal vectors 
\begin{equation}
\label{v2}
|v^{{}_{(2)}}_{m}\rangle\equiv [\widetilde N_2]^m_{\nu_{2L},\nu_{2R}}
|w^{{}_{(2)}}_{\nu_{2L},
\nu_{2R}}\rangle. 
\end{equation}
Using~\eqref{v1} and~\eqref{v2} one can rewrite $\rho_{A_1\cup A_2}$ as 
\begin{multline}
\label{rho-fin}
\rho_{A_1\cup A_2}=\\
\big([N_1]^{m}_{\nu_{1L},\nu_{1R}}[\bar N_1]^{m'}_{\nu_{1L},\nu'_{1R}}\big)
\big([N_2]^{n}_{\nu_{2L},\nu_{2R}}
[\bar N_2]^{n'}_{\nu'_{2L},\nu_{2R}}\big)\\
\times[E_{12}]^{\nu_{1R},\nu_{2L}}_{\nu'_{1R},\nu'_{2L}}
\,\,|v^{{}_{(1)}}_m\rangle\otimes|v^{{}_{(2)}}_n\rangle\langle 
v^{{}_{(1)}}_{m'}|\otimes\langle v_{n'}^{{}_{(2)}}|, 
\end{multline}
where $N_1\equiv \widetilde{N}_1^{-1}=UD^{\frac{1}{2}}$, and similarly for $N_2$. 
The definition of $N_1$ from the SVD of $E_1$ is illustrated in Fig.~\ref{cartoon2}. 
Formula~\eqref{rho-fin} is illustrated pictorially in Fig.~\ref{cartoon3} (a). 
In~\eqref{rho-fin} all the indices are contracted except for $m,m',n,n'$. Precisely, 
the index pair $(m,m')$ refers to subsystem $A_1$, whereas $(n,n')$ is associated 
with $A_2$. Moreover, one has that $n,m,n',m'\in [1,\chi^2_{max}]$, implying that 
the computational cost to evaluate the terms in the round brackets in~\eqref{rho-fin} 
is $\propto\chi_{max}^7$. However, this can be reduced by considering a truncated SVD 
of $E_1$ and $E_2$, i.e., keeping only the $p$ ($p\ll\chi_{max}^2$) largest singular 
values. This relies on the fact that the singular values of $E_1$ and $E_2$ typically 
decay very quickly. The computational cost is now $p^2\chi_{max}^3$. The same trick 
is used in the implementation of DMRG with periodic boundary conditions~\cite{pippan-2010}. 
Finally, from~\eqref{rho-fin} it is straightforward to obtain $\rho_{A_1\cup A_2}^{T_2}$ 
by exchanging $n\leftrightarrow n'$, as illustrated in Fig.~\ref{cartoon3} (b). 

\subsection{The clean $XX$ chain: Two adjacent intervals} 
\label{clean-dmrg}


\begin{figure}[t]
\includegraphics*[width=0.9\linewidth]{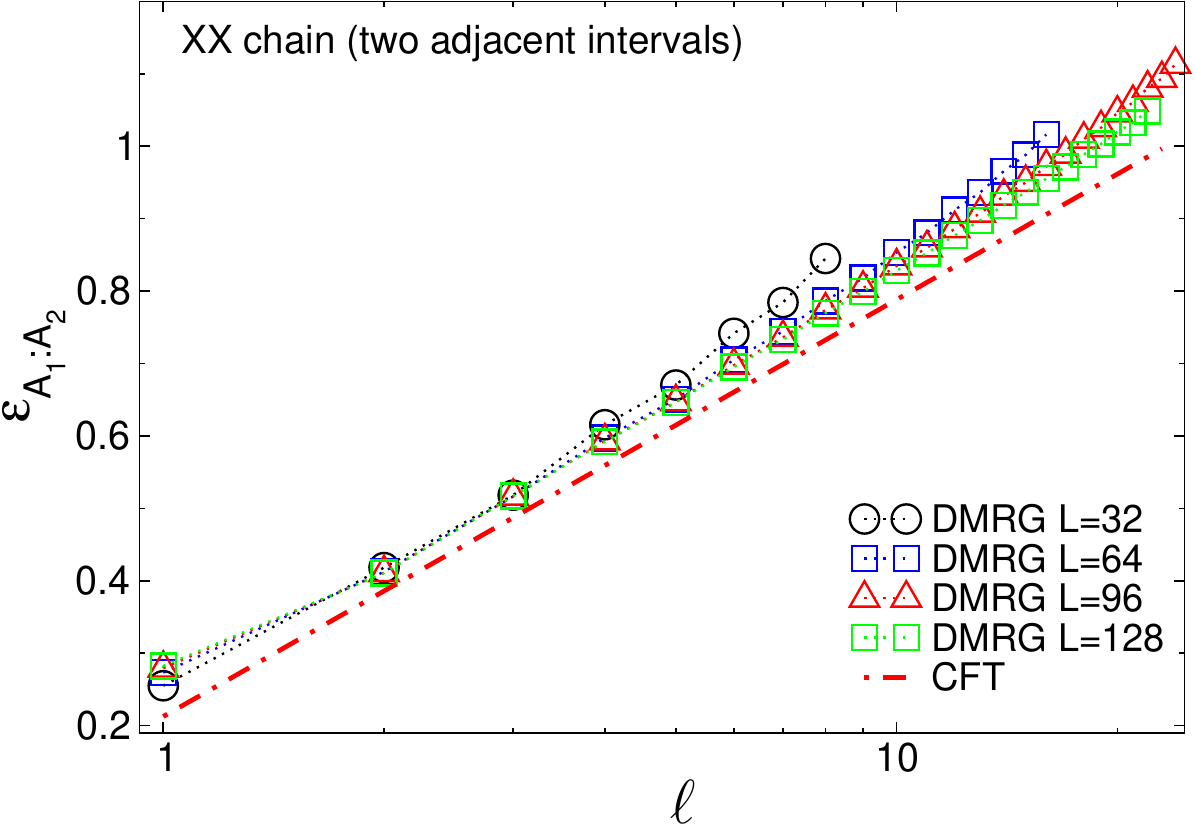}
\caption{ The logarithmic negativity ${\cal E}_{A_1:A_2}$ between the two 
 (equal-length) adjacent intervals $A_1\cup A_2$ in the $XX$ chain. Here the 
 two intervals are at the center of the chain. ${\cal E}_{A_1:A_2}$ is plotted 
 against the length of one interval $\ell$. The symbols are DMRG data for 
 a chain with length up to $L=128$. The dash-dotted line is the CFT prediction 
 $1/4\ln\ell$ in the thermodynamic limit. 
}
\label{clean-fig2}
\end{figure}

Before moving to disordered systems it is worth providing some checks of~\eqref{rho-fin} 
for clean systems in which all results are under control. To this purpose here we  focus 
on the clean  $XX$ chain 
\begin{equation}
{\mathcal H}_{XX}=J\sum\limits_{i=1}^{L-1}(S^x_iS^x_{i+1}+S^y_i
S^y_{i+1})+h\sum\limits_{i=1}^{L}S_i^z, 
\label{xx_ham1}
\end{equation}
with open boundary conditions. 
We fix $J=1$, considering zero magnetic field $h=0$. We restrict ourselves to 
two adjacent intervals at the center of the chain (see Fig.~\ref{figure2} (a)). 
This corresponds to the trivial transfer matrix $[E_{12}]^{\nu_{1R},\nu_{2L}}_{
\nu'_{1R},\nu_{2L}'}=\delta_{\nu_{1R}\nu_{2L}}\delta_{\nu'_{1R}\nu'_{2L}}$ 
in~\eqref{rho-fin}.

DMRG data for the logarithmic negativity ${\cal E}_{A_1:A_2}$ between  two adjacent 
intervals of equal length $\ell$ are reported in Fig.~\ref{clean-fig2}. 
The different symbols correspond to $L=32,64,96,128$. In the DMRG simulations we fixed the largest bond dimension 
to $\chi_{max}=80$, and the discarded weight~\cite{uli2} to $\epsilon\sim 10^{-11}$. Also, 
to reduce the computational cost, we considered a truncated SVD of the transfer 
matrices $E_1$ and $E_2$ (cf.~\eqref{rho-fin} and Fig.~\ref{cartoon2}) 
fixing $p=200$.


The DMRG data for ${\cal E}_{A_1:A_2}$ are plotted against the length of the two intervals $\ell$ 
(notice the logarithmic scale on the $x$-axis). 
There are no CFT predictions for ${\cal E}_{A_1:A_2}$ in finite systems with open boundary conditions,  
so we can only use the result for two adjacent intervals in the infinite chain~\cite{calabrese-2012}
\begin{equation}
\label{adj-cft}
{\cal E}_{A_1:A_2}=\frac{c}{4}\ln\Big(\frac{\ell_1\ell_2}{\ell_1+\ell_2} \Big)+\textrm{cnst},  
\end{equation}
with $\ell_1$ and $\ell_2$ the lengths of the two intervals, $c$ the central 
charge, and $\textrm{cnst}$ an additive constant. 
This is expected to describe the data for an open chain as long as $\ell_1,\ell_2\ll L$. 
This is confirmed for the $XX$ chain by the data in Fig.~\ref{clean-fig2} in which 
the dash-dotted line is~\eqref{adj-cft} for $\ell_1=\ell_2=\ell$. 
While some deviations from~\eqref{adj-cft}  are visible for finite $L,\ell$, the agreement with the CFT prediction~\eqref{adj-cft} 
becomes progressively better upon increasing $L$, in the region $\ell\ll L$. 

\section{Negativity in the random $XX$ chain: Some DMRG results}
\label{neg-dmrg1}

Here we discuss the logarithmic negativity ${\cal E}_{A_1:A_2}$ of two adjacent intervals 
in the random $XX$ chain ($\Delta=0$ in~\eqref{xxz-ham}, see Appendix~\eqref{dis-XX}). 
To this purpose we perform DMRG simulations, focusing on the disorder distribution $P_\delta(J)=\delta^{-1}J^{-1+1/\delta}$, 
with $\delta=1/4,1,3/2$. Increasing values of $\delta$ correspond to increasing disorder strength. 
Specifically, $\delta=0$ corresponds to the clean case, whereas for $\delta\to\infty$ one recovers the infinite-randomness 
fixed point (IRFP) distribution (cf.~\eqref{Pfixedpoint}). Our data for ${\cal E}_{A_1\cup A_2}$ correspond 
typically to an average over $\sim 10^3$ disorder realizations. 
Although exact results for the negativity are not yet available, many entanglement-related quantities, such as the 
entanglement entropy, can be calculated  for the random $XX$ chain, via a mapping 
to free fermions (cf. Appendix~\ref{dis-XX}). Importantly,  this provides 
a reliable way to check the accuracy of DMRG results.

\subsection{Convergence of DMRG}

For each disorder realization, a crucial aspect is the convergence of the DMRG 
method. This depends both on the chain sizes $L$ and on the disorder strength 
$\delta$. Specifically, we numerically observed that the convergence of DMRG 
becomes rapidly poor upon increasing $L$ or $\delta$. This is physically 
expected because the disorder gives rise to a ``rough'' energy landscape, 
which makes likely for DMRG to get trapped in a local minimum. This affects  
severely the convergence because typical DMRG update schemes are {\it local}~\cite{uli2}. 
This is also related to the exponential small gap~\cite{Fisher1994} $\ln 
\Delta E \sim -L^{1/2}$ of the RS phases. 

To check the correctness of the DMRG results, for each disorder realization we compared 
the data for the half-chain von Neumann entropy with the exact calculation using 
free-fermion techniques (see Appendix~\ref{dis-XX}). We used as a DMRG convergence 
criterion the matching of the two results within a precision of $10^{-3}$. We also defined 
the DMRG convergence rate $R_c$ as the fraction of converged DMRG simulations. Only converged 
disorder configurations were included in the disorder averages. 

The fact that $R_c<1$ results in a systematic error in the DMRG data. This can be mitigated 
by using a large number of DMRG sweeps. We typically used $n_{sw}\approx 100$ in our simulations. 
Moreover, we observed that in each sweep it is crucial to perform a sufficient 
number of Lanczos iterations $n_{iter}$ when locally optimizing the MPS. In our simulations we 
used $n_{iter}\approx 100$. Still, we should stress that this is not sufficient to ensure 
$R_c=1$. For instance, we observed that for $L=63$ at $\delta=1$, using $n_{sw}=100$ and 
$n_{iter}=100$, one has $R_c\approx 0.9$. Finally, to provide reliable error bars for our 
DMRG results, we always checked that the systematic error was negligible compared 
with the statistical error arising from the disorder average, providing the latter as 
our final error estimate. 

\begin{figure}[t]
\includegraphics[width=0.88\linewidth]{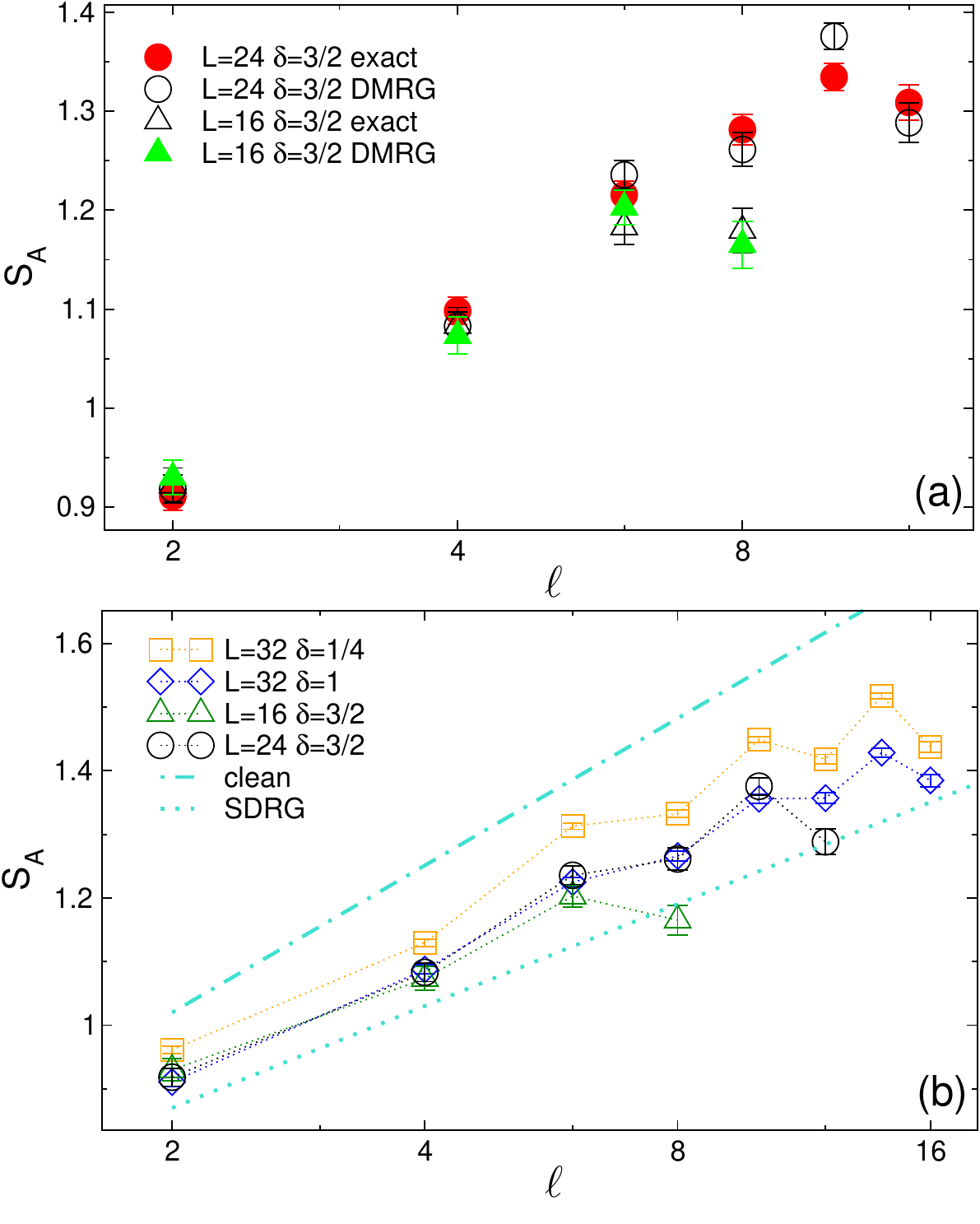}
\caption{Entanglement entropy in the random XX spin chain.  
 Panel (a): Check of the DMRG convergence. The von Neumann entropy $S_{A}$ for 
 a single interval at the center of an open  chain plotted versus the interval length $\ell$. 
 The empty symbols denote DMRG results for chains with $L=16,24$ and disorder 
 strength $\delta=3/2$. The data are averaged over $\sim 10^3$ disorder 
 realizations. The full symbols are exact results obtained using free-fermion 
 techniques. Panel (b): Same geometry as in (a). The symbols are DMRG data for 
 $L=16,24,28,32$ and $1/4\le\delta \le 3/2$. Here $\delta=0$ and $\delta\to\infty$ 
 correspond to the clean case and the IRFP, respectively. For $\delta=1$ one has the 
 box distribution. The dash-dotted line is the CFT prediction for $\delta=0$. The dotted 
 line is the strong disorder renormalization group (SDRG) prediction in the thermodynamic 
 limit. 
}
\label{dmrg2}
\end{figure}

Fig.~\ref{dmrg2} (a) shows the comparison between the DMRG data for $S_{A}$ (empty symbols 
in the figure) and the exact results (full symbols) obtained using free-fermion methods 
(see Appendix~\ref{dis-XX}). Data are for $\delta=3/2$, which is the most difficult 
value of $\delta$ to simulate, and $L=16,24$. The error bars are the statistical 
errors resulting from the disorder average. In all cases the DMRG results are in agreement 
with the free-fermion result, within error bars. This suggests that, at least 
for the von Neumann entropy, the systematic error due to $R_c<1$ is 
negligible compared to the statistical one. 

\subsection{DMRG results}

Fig.~\ref{dmrg2} (b) reports 
the von Neumann entropy $S_{A}$ for the interval $A$ 
at the center of the open chain. 
The symbols are DMRG data for different $\delta$ and $L$. The dash-dotted 
line is the CFT result in the clean case $S_{A}\simeq 1/3\ln\ell+c'$, 
whereas the dotted line is the SDRG result 
\begin{equation}
\label{pred}
S_{A}\simeq\frac{\ln 2}{3}\ln \ell+{\rm const}. 
\end{equation}
We recall that the fixed point SDRG is expected to describe the numerical data for any disorder strength, 
but for asymptotically large $\ell$ and $L$. 
For finite interval and system lengths, we expect  a complicated crossover between the clean and SDRG results, 
which is more severe for small disorder. 
In fact, from Fig.~\ref{dmrg2} (b) it is clear that for the smallest disorder strength ($\delta=1/4$),
the data are closer to the clean prediction than to the SDRG one. 
This does not come unexpected since the crossover between the two fixed points 
should be regulated by a $\delta$-dependent crossover length  $\xi_\delta$. 
One should expect that $\xi_\delta\to\infty$ for $\delta\to0$. 
At any $\delta$, the IRFP results should be valid only for $\ell,L\gg\xi_\delta$. 
However,  the data for $\delta=1$ and $\delta=3/2$ are almost indistinguishable, signaling that 
they should be both in the SDRG regime.  
Indeed, although oscillating corrections are present (as it is well known~\cite{FagottiCalabreseMoore2011}), 
the data are in rough agreement with the SDRG prediction~\eqref{pred} (dotted line). 
A more quantitative analysis of the entanglement entropy, with a proper robust determination of the prefactor to 
the logarithmic growth, requires the study of much longer chains as done by means of  
free fermionic methods\cite{laflorencie-2005,FagottiCalabreseMoore2011}.  
Unfortunately, simulating with DMRG chains of length of the order of hundreds sites and with sufficient statistics is beyond 
our current capability. 

\begin{figure}[t]
\includegraphics[width=0.9\linewidth]{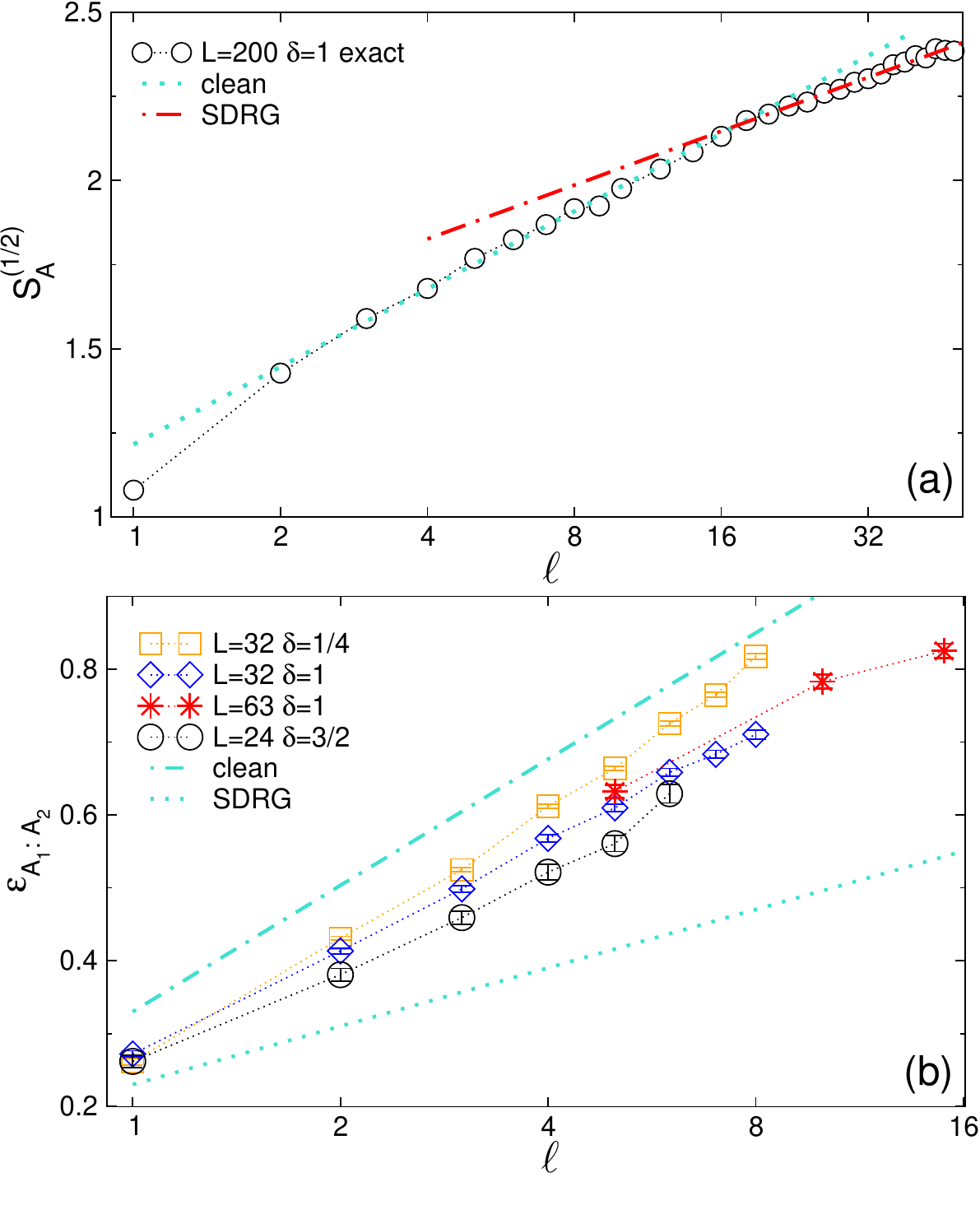}
\caption{Entanglement negativity in the random XX spin chain.  
 Panel (a): R\'enyi entropy $S^{(1/2)}_A$ for a single interval  at the center of an open  chain plotted versus the interval length $\ell$.
 The data (circles) are obtained by means of free-fermionic techniques which allow us to reach large systems sizes ($L=200$).  
 Note the crossover from the scaling of the clean model at short lengths (dotted line) to the SDRG one at large distances 
 (dash-dotted line). 
 Panel (b): The logarithmic negativity ${\cal E}_{A_1:A_2}$ for two adjacent intervals of equal length $\ell$
 (in the center of an open chain) plotted as a function of $\ell$. 
 The symbols denote DMRG results for chains with $L=16,24,28,32,63$ and disorder strength $1/4\le\delta\le3/2$. 
 The data are averaged over $\sim 10^3$ disorder realizations.
 The dash-dotted and dotted lines are the CFT prediction for $\delta=0$, and the SDRG result, respectively. 
}
\label{dmrg3}
\end{figure}

Although the data for the entanglement entropy are asymptotically in rough agreement with the prediction of the 
SDRG already for $L=32$ and $\delta\sim1$, this does not imply that the same is true for the negativity,
because the crossover between the CFT (clean) and the IRFP fixed points in principle depends on the measured 
quantity. In order to understand the scaling of the negativity, it is worth to consider the R\'enyi entropy 
\begin{equation}
S^{(1/2)}_A\equiv 2 {\rm Tr} \rho_A^{1/2},
\end{equation} 
for a single interval $A$ of length $\ell$ at the center of an open chain.
Indeed, the negativity for pure states (e.g., for two adjacent intervals with $\ell_1+\ell_2=L$)  
coincides with  $S^{{}_{(1/2)}}_{A}$. 
Consequently, one could expect the scaling of the negativity to resemble that of $S^{{}_{(1/2)}}_A$, 
rather than that of the von Neumann entropy $S_A$.
In Fig. \ref{dmrg3} (a) we report exact numerical data for $S_A^{{}_{(1/2)}}$ at $\delta=1$ obtained by means of 
free-fermionic techniques, which allow us to reach large systems sizes ($L=200$).
Interestingly, the data do not show strong parity effects (i.e., oscillations with  
the parity of the block size), in contrast with $S_{A}$ (the same 
is true for the negativity in Fig. \ref{dmrg3} (b)). 
Considering these large system sizes we can clearly see the crossover from the scaling of the clean model 
 at short lengths ($S^{{}_{(1/2)}}_A\sim 1/2 \ln \ell$) to the SDRG one at large distances 
($S^{{}_{(1/2)}}_A\sim (\ln 2)/3 \ln \ell$). The crossover starts around $\ell\sim 15$ and 
it is fully established around $\ell\sim 30$, implying that $L\approx 120=4\times 30$ would be 
needed to fully confirm our results. This, however, is not possible with our current DMRG implementation.

Let us, however, consider the logarithmic negativity ${\cal E}_{A_1:A_2}$ between the 
two adjacent intervals. The DMRG data are reported in Fig.~\ref{dmrg3} (b) for chain sizes up to $L=63$ and 
for several disorder strengths.   
The dash-dotted line is the CFT prediction for the clean model ${\cal E}_{A_1:A_2}\simeq 1/4\ln(\ell)+c'$ (cf. Fig.~\ref{clean-fig2}) 
while the dotted line is the SDRG result ${\cal E}_{A_1:A_2}\simeq(\ln 2)/6\ln(\ell)+ {\rm const}$ (cf.~\eqref{neg_adj}). 
For $\delta=1/4$, ${\cal E}_{A_1\cup A_2}$ grows with a  slope similar to the clean case 
(compare also with Fig.~\ref{clean-fig2}). 
This does not come unexpected since also the results for the entanglement entropy in Fig. \ref{dmrg2} (b) 
for $\delta=1/4$ are affected by a strong crossover. 
Increasing the strength of the disorder, we observe that for the available subsystems' sizes 
the slopes of the negativity appears to reduce, but, as a difference with the entanglement entropy, 
there is no saturation with $\delta$ and also for $\delta=3/2$ there are still strong crossover effects. 
This is in complete analogy with what observed for $S^{(1/2)}_A$ in Fig. \ref{dmrg3} (a), where the 
crossover for $\delta=1$ starts taking place around $\ell=15$. 
Indeed, for $\delta=1$, the last points for $L=63$ with $\ell=15$ seem to move in the right direction 
(we cannot access chains of this length for $\delta=3/2$ because of the poor convergence of DMRG for this 
large disorder). However, these are only very encouraging signals for the correctness of our prediction: 
we can only conclude that for the system sizes and $\delta$ accessible to DRMG simulations, 
${\cal E}_{A_1:A_2}$ exhibits deviations from the SDRG prediction~\eqref{neg_adj}, which are likely explained 
as crossover effects.  
Much larger chain sizes (of the order of $L\sim120/130$ which are not accessible with our current DMRG implementation), 
would be needed for the asymptotic IRFP behavior to set in.

\section{Discussion}
\label{concl}

In this manuscript we considered the scaling behavior of the logarithmic negativity of disordered spin chains 
in the random singlet phase. 
Our main results have been already summarized in the Introduction, and for this reason here we only mention 
some new research directions originating from the present work. 
First, in the random $XX$ chain it should be possible by generalizing the results for clean 
systems~\cite{eisler-2014,coser-2015,ctc-16}  to treat the moments of the partially-transposed 
density matrix exactly. By comparing with the analytic results of section~\ref{SECTIONMomentsRSP}, 
this would allow to provide a more robust check of some of our findings. 
Another  intriguing direction would be to investigate to which extent the relation between the negativity 
and the mutual information remains valid in more complicated disordered phases. In this 
respect one possibility would be to focus on the spin-$1$ random Heisenberg 
chain~\cite{refael-2007}. 
Different physical behavior should also appear in the spin-$1/2$ Heisenberg chain in which, besides antiferromagnetic 
couplings, ferromagnetic ones are allowed. 
Finally, it would be interesting to investigate the scaling of the negativity 
in excited states of disordered spin models, also in connection with many-boby localization.

\section{Acknowledgments}

We acknowledge very useful discussions with G.~Sierra and K.~Agarwal. We also 
would like to thank M.~Collura, F.~Pollmann and M.~Stoudenmire for very useful insights 
on the DMRG simulations, and on the ITensor library. All the authors acknowledge support 
from the ERC under the Starting Grant 279391 EDEQS.  

\appendix 

\section{The disordered $XX$ chain}
\label{dis-XX}

The random $XX$ chain with open boundary conditions is defined by the 
Hamiltonian 
\begin{equation}
{\mathcal H}_{XX}=\sum\limits_{i=1}^{L-1}J_i(S^x_iS^x_{i+1}+S^y_i
S^y_{i+1})+h\sum\limits_{i=1}^{L}S_i^z, 
\label{xx_ham}
\end{equation}
with $S^{x,y,z}_i\equiv\sigma_i^{x,y,z}/2$, $\sigma_i^\alpha$ 
being the Pauli matrices acting on site $i$. For periodic boundary conditions 
one has an extra term in Eq.~\eqref{xx_ham} connecting site $L$ with site $1$. 
Hereafter we fix $h=0$. Here $J_i$ are uncorrelated random variables. After 
the Jordan-Wigner transformation 
\begin{equation}
c_i=\Big(\prod\limits_{m=1}^{i-1}\sigma^z_m\Big)
\frac{\sigma_i^x-i\sigma_i^y}{2},
\label{j-wigner}
\end{equation}
\eqref{xx_ham} is recast in the free-fermionic form 
\begin{equation}
{\mathcal H}_{XX}=\frac{1}{2}\sum\limits_{i=1}^{L-1}J_i(c^\dagger_i 
c_{i+1}+c^\dagger_{i+1}c_i)+\frac{h}{2}\sum\limits_{i=1}^{L-1}
c^\dagger_i c_i,
\label{xx_fer}
\end{equation}
with $c_i$ spinless fermionic operators satisfying the canonical 
anticommutation relations $\{c_m,c^\dagger_n\}=\delta_{m,n}$. 

One can now impose that the single-particle eigenstates $|\Psi_q\rangle$ (with $q$ 
an integer labelling the different eigenstates) of~\eqref{xx_fer} are of 
the form 
\begin{equation}
\eta_q^\dagger|0\rangle\equiv|\Psi_q\rangle=\sum_i\Phi_q(i)c_i^\dagger|0\rangle,
\end{equation}
with $|0\rangle$ the fermion vacuum, $\Phi_q(i)$ the eigenstate amplitudes, 
and $\eta_q$ new fermionic operators. The Schr\"odinger equation, which 
allows to determine $\Phi_q(i)$, becomes  
\begin{equation}
\label{xx-eig}
J_i\Phi_q(i+1)+J_{i-1}\Phi_q(i-1)=2\epsilon_q\Phi_q(i),\quad\forall i, 
\end{equation}
with $J_L=0$, and $\epsilon_q$ the single-particle eigenvalues. Notice 
that~\eqref{xx-eig} is the eigenvalue problem for the banded matrix $T
\equiv(J_j\delta_{i,j+1}+J_{j-1}\delta_{i,j-1})/2$. One can show that 
the eigenvalues of the matrix $T$ are organized in pairs with opposite sign. 
Precisely, given the amplitude $\Phi_1(i)$ of an eigenvector with $\epsilon_q
>0$, the amplitudes of the eigenvector with eigenvalue $-\epsilon_q$ are 
obtained as $(-1)^{i+1}\Phi_q(i)$.

As a consequence, the ground state $|GS\rangle$ of~\eqref{xx_ham} is 
in the sector with $M=L/2$ fermions. This is constructed by filling 
all the negative modes $\epsilon_q<0$ as 
\begin{equation}
|GS\rangle=\eta^\dagger_{q_M}\eta^\dagger_{q_{M-1}}\cdots\eta_{
q_1}^\dagger|0\rangle.
\end{equation}
It is straightforward to derive the anticommutation relations
\begin{equation}
\label{us1}
\{\eta^\dagger_q,c^\dagger_j\}=\{\eta_q,c_j\}=0, 
\end{equation}
and
\begin{equation}
\label{us2}
\{\eta_q^\dagger,c_j\}=\Phi_q(j)\delta_{k,j},\quad\{\eta_q,
c^\dagger_j\}=\Phi^*(j)\delta_{k,j}. 
\end{equation}
Using~\eqref{us1} and~\eqref{us2}, the expectation value of the 
two-point function $\langle c^\dagger_ic_i\rangle$ in a generic 
eigenstate of~\eqref{xx_ham} is given as 
\begin{equation}
\label{corr-f}
\langle c_i^\dagger c_j\rangle=\sum_{q}\Phi_q^*(i)\Phi_q(j), 
\end{equation}
where the sum if over the $q$ single-particle excitations forming 
the eigenstate.

In order to calculate the entanglemen entropy, let us now introduce a bipartition of the chain into two subsystems as 
$A\cup\bar A$, with $\bar A$ denoting the complement of $A$. For free-fermion 
models, even in presence of disorder, the reduced density matrix $\rho_A$ of 
subsystem $A$ is determined by the correlation matrix~\cite{peschel-1999,
peschel-1999a,chung-2001,peschel-2004,peschel-2004a,peschel-2009} restricted to 
$A$ as 
\begin{equation}
{\mathcal C}^{(A)}_{ij}\equiv\langle c_i^\dagger c_j\rangle, 
\end{equation}
where $i,j\in A$. In particular, given the eigenvalues $\lambda_k$ of 
${\mathcal C}^{{(A)}}$, the entanglement entropy $S_A$ is given as 
\begin{equation}
\label{sa-ff}
S_A=-\sum_k(\lambda_k\ln\lambda_k+(1-\lambda_k)\ln(1-\lambda_k)).
\end{equation}
%



\end{document}